\documentclass[final,3p,times,numbers]{elsarticle}%,numbers
\usepackage{hyperref}%lineno,
\usepackage{amsmath,amssymb}
\usepackage{caption}
\usepackage{subcaption}
\biboptions{numbers,sort&compress}
\journal{}%Journal of \LaTeX\ Templates
\begin{document}

\begin{frontmatter}

\title{Analyzing the equilibrium states of quasi-neutral spatially inhomogeneous system of charges above liquid dielectric film basing on first principles of quantum statistics%\tnoteref{mytitlenote}
	}

%[mycorrespondingauthor]

\author{Yu.V.~Slyusarenko\corref{coruv}}

\author{D.M.~Lytvynenko %\fnref{myfootnote}
	\corref{mycorrespondingauthor}}

%\author[mymainaddress,mysecondaryaddress]{Elsevier Inc}\ead[url]{www.elsevier.com}

%\author[mysecondaryaddress]{Global Customer Service}

%[mymainaddress]
\address{NSC KIPT, Akhiezer Institute for Theoretical Physics, Akademichna Street 1, Kharkiv 61108, Ukraine}
%[mysecondaryaddress]
\address{V.N.~Karazin Kharkiv National University, High-Technology Institute, Kurchatov Avenue 31, Kharkiv 61108, Ukraine}

\cortext[mycorrespondingauthor]{Corresponding author}
\ead{d\_m\_litvinenko@kipt.kharkov.ua}

\begin{abstract}
The theory of quasi-neutral equilibrium states of charges above liquid dielectric surface is built. This theory is based on first principles of quantum statistics for systems, comprising many identical particles.
The proposed approach is concerned with applying the variational principle, modified for the considered systems, and the Thomas-Fermi model.
In terms of the developed theory a self-consistency equations are obtained. These equations provide the relation between the main parameters, describing the system: the potential of static electric field, the distribution function of charges and the surface profile of liquid dielectric.
The equations are used to study the phase transition in the system to a spatially periodic state.
The proposed method can be applied to analyzing the properties of the phase transition in the system to a spatially periodic states of wave type.
Using the analytical and numerical methods, we make a detailed research of the dependence of critical parameters of such phase transition on the thickness of liquid dielectric film.
Some stability criteria of the new asymmetric phase of the studied system are discussed.
\end{abstract}


\end{frontmatter}

%\linenumbers

\section{Introduction}
\paragraph{}
In spite of more than forty-year history~\cite{pla1971CrandallW,jetpl1974Shikin,jetp1975MonarkhaS,prl1979FisherHP} the research of phenomena, concerned with the formation of spatially periodic states in a system of charged particles above dielectric surface, is still relevant.
The possibility of spatially periodic ordering in three-dimensional systems of charges (electrons in metals) was predicted by Wigner~\cite{pr1934Wigner}.
Due to this paper one can meet the ``Wigner crystallization'' term in scientific literature.
Many years after it was shown, that the phase transition in such system to a three-dimensional spatially periodic state formation could be predicted in other way~\cite{ltp1998PPS}.
For decades the experimental realization of WC in three-dimensional systems failed.
However, the stable spatially periodic states of charges were experimentally realized in the system of electrons near the boundary, separating two media.
The examples of such systems are: electrons above the surface of liquid helium film in the external clamping electric field~\cite{prl1979GrimesA, pla1979LeidererW}, the electrons on the surface of solid hydrogen and neon~\cite{jetpl1979TroyanovskiiVH, ss1984Kajita}, as well as the system of electrons at the interface between the semiconductor p-n junction~\cite{prb1982BishopDT}.
The detection of stable systems, convenient for experimental research of spatially periodic structures of charges near media interfaces, greatly increased the interest in this kind of research.
Chronology of the research can be followed in the books~\cite{book1997Andrei, book2003MonarkhaK} and review articles~\cite{ltp1982MonarkhaS, ltp2012MonarkhaS, ufn2011Shikin}.

The available works describe the effects, concerned with as two-dimensional (2D) Wigner crystals~\cite{prl1979GrimesA}, as macroscopic dimple lattices~\cite{pla1979LeidererW}.
The theoretical works, describing the corresponding experiments, are usually based on the concept of the energy spectrum of a single (or ``levitating'') electron above the dielectric surface.
This concept considers a single electron, located above flat dielectric surface, with its electrostatic image  as an analogue of a hydrogen atom with the corresponding energy spectrum~\cite{prl1969ColeC}.
Obviously, in the case of many particle system, such approach faces not only the mathematical difficulties, but also the ``philosophical'' or methodological ones.
Let us remind, that the electrostatic image method can be treated only as a mathematical ``trick'' in the case of a single charge, that is ``fixed'' above the metal or dielectric surface. This method allows avoiding the consecutive solving of the Poisson equation.
The mentioned difficulties vanish in the case of microscopic theory of the system description.
Such theory should consider the researched system as a quantum mechanical many-body system~\cite{jmp2012LytvynenkoSK, past2012SlyusarenkoSK, cmp2009LytvynenkoS, jps2015SlyusarenkoL} and take into account the external electric clamping field.
This field plays an important role in the formation of such systems, because the attracting field, generated by charges in dielectric, is insufficient for holding them near the surface.

The basics of such microscopic theory development were formulated in Ref.~\cite{jmp2012LytvynenkoSK}.
The method is based on the variation principle, modified for the considered system, and the Thomas-Fermi model.
This variation principle takes into account the external electrostatic clamping field.
This approach provides obtaining the self-consistent equations, relating the parameters of the system description (the potential of electric field, the  distribution function of charges and the surface profile of liquid dielectric).
As an application of the developed theory the authors studied phase transitions to a spatially periodic states in the system of charges above the surface of liquid dielectric.
The approach to solving the self-consistency equation system was outlined and the parameters of the phase transition were obtained. Besides that, the period of the reciprocal lattice of spatially periodic structures  was obtained too.
As an approbation of the developed theory, its results were compared to the experimental data~\cite{pla1979LeidererW}.
As the result, the qualitative agreement of the theoretical and experimental data were obtained.

However, let us emphasize the essential feature of Ref.~\cite{jmp2012LytvynenkoSK}.
This paper was devoted to the describing of the system with fixed number of charges above dielectric surface.
So as Ref.~\cite{cmp2009LytvynenkoS} was.
In other words, the system was considered to be charged, but not quasi-neutral.
Traditionally a system is considered to be quasi-neutral, if its number of charges is not fixed and it is defined by the external field.
In this case the number of charges is exactly as needed to compensate  external electrostatic clamping field.
For this reason, the electric field vanishes at large distance from the dielectric surface~\cite{book1997Andrei,book2003MonarkhaK,ltp1982MonarkhaS,ltp2012MonarkhaS,ufn2011Shikin}.
In the experimental research both system types are used.
However, quasi-neutral system of charges above liquid dielectric surface has a number of significant features~\cite{prl1979GrimesA,book2003MonarkhaK} in comparison with  ``charged'' systems, studied in Refs.~\cite{jmp2012LytvynenkoSK,cmp2009LytvynenkoS}.
E.g., the spatial distribution of charges and electric fields can be  considerably different in these two system types.
Some of these features can be predicted only if the system is described in terms of microscopic theory~\cite{jps2015SlyusarenkoL}.
The above circumstances are the main motivation for writing this paper, which is dedicated to a consequent statistical description of the equilibrium state of the quasi-neutral system of charges above the surface of liquid dielectric film.
Like in the case of charged systems~\cite{jmp2012LytvynenkoSK, cmp2009LytvynenkoS}, the capabilities, provided by the theory, proposed in this paper, are demonstrated on the study of the phase transition in quasi-neutral systems to the state with a spatially periodic ordering.
Special attention in this research is given to the analysis of the influence of the thickness of liquid dielectric film on the physical features of such phase transitions.
As the limit cases the ``thick'' and ``thin'' helium films are taken.
In the case of thick helium films (e.g., the so-called bulk helium~\cite{pb1984Leiderer,ss1982LeidererES}) the main contribution to the force, acting on the unit of liquid dielectric volume, is made by the gravitational attraction.
On the other hand, in the case of thin helium films the gravitational attraction is negligibly small comparing to Van der Waals interaction forces between helium atoms and the substrate substance~\cite{prb1981IkeziP,prb1990HuD}.
The results, obtained in this paper, are compared with the existing data, obtained in other description models and experimental studies.

The basics of the proposed theory can be used for the description of other systems, such as heavy ions in gravitational field above dielectric surface.
The above mentioned system can serve as a model for the research of spatial distribution of ``levitating'' radiation dust above dielectric surfaces.
This fact is relevant for ecological research, concerned with Chernobyl problems, e.g.

It should be also noted, that the statistical approach to the description of spatially inhomogeneous states in systems of particles with Coulomb interaction (including the electronic system on the surface of liquid helium) was used in Refs.~\cite{pre1998LevZ,pre2011LevZ,ujp2015LevOTZ,epjb2014LevOTZ}.
The methodology of these studies was based on using the modified electrostatic potential of a single electron and methods of functional integration to calculate the grand partition functions.

\section{Self-consistency equations for the system of charges above liquid dielectric surface}

Unlikely the charged system, in the quasi-neutral one the total number of charges is determined by the external field.
In other words, if the external clamping field changes, the number of charges, held above the dielectric surface, changes too.
However, the difference between the system types does not affect the basic formulations of the statistical approach to their description~\cite{jps2015SlyusarenkoL}.
The difference between the quasi-neutral and charged systems appears at the level of boundary conditions formulation for the self-consistency equation system.
For this reason, this article does not contain a detailed description of the theory basics.
Following Ref.~\cite{jps2015SlyusarenkoL} in the present article, we briefly annotate the terminology and formulations, resulting in  self-consistent equations.

\begin{figure}
	\centering
	\includegraphics[scale=0.5]{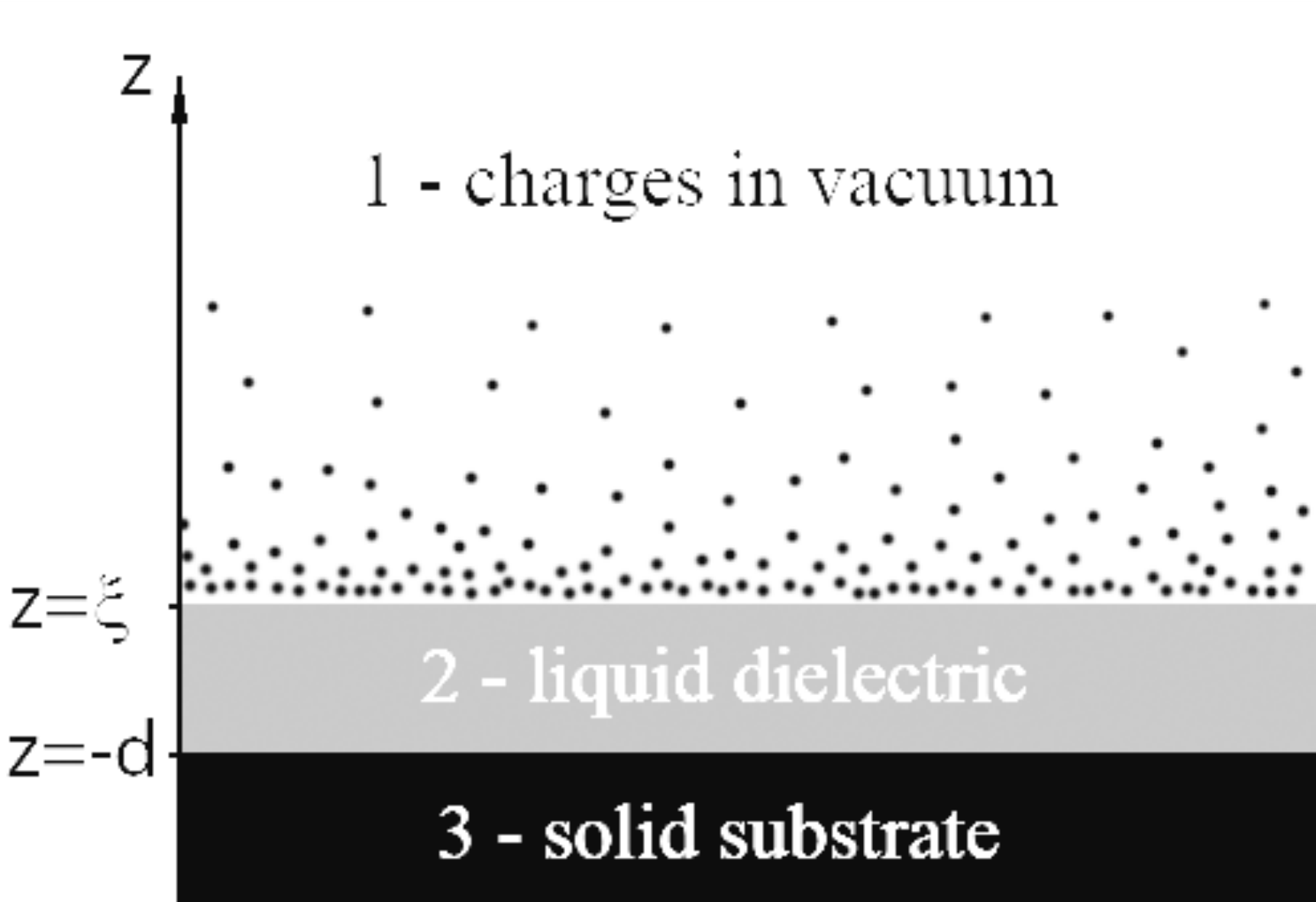}
	\caption{\label{fig:epsart}  System of charges in region~``1'' above liquid dielectric~``2'' surface  on  solid substrate~``3''.}
\end{figure}

Let us consider a system of identical particles, having charge $Q$, mass $m$, spin $S_Q$, momentum ${\bf{p}}$ and energy ${\varepsilon _{\bf{p}}} = \frac{{{{\bf{p}}^2}}}{{2m}}$.
The charges are placed in vacuum above the surface of liquid dielectric film, having thickness $d$. The liquid dielectric has dielectric permittivity $\varepsilon$ and surface tension coefficient $\alpha$.
We assume, that the liquid dielectric film is located on a flat solid substrate, having dielectric permittivity ${\varepsilon _d} \gg \varepsilon$.
The surface profile of liquid dielectric film is described by $\xi ({\boldsymbol{\rho }}) \equiv \xi (x,y)$ function, where ${\boldsymbol{\rho }} \equiv \left\{ {x,y} \right\}$ is the radius vector in $z = 0$ plane of the Cartesian coordinate system $\{ z,x,y\} $.
The boundaries between ``1'' - ``3'' regions  (see Fig.~\ref{fig:epsart}) in the direction of ${\boldsymbol{\rho }}$ coordinates are considered unlimited.
To avoid issues, concerned with repulsion of like-charged particles along ${\boldsymbol{\rho }}$, the system is assumed to be located in a vessel with walls at $\rho  \to \infty $.
These walls forbid charges to leave the system  along the undisturbed flat surface of liquid dielectric.

The charges are acted by external clamping electric field $E_1^{\left( e \right)}$, directed along $z$-axis.
It is also assumed, that there is a potential barrier prohibiting the charges entry into the liquid dielectric film.
All physical quantities related to $z > \xi ({\boldsymbol{\rho }})$ region are marked by ``1''~index, the physical quantities, related to liquid dielectric  film ($\xi ({\boldsymbol{\rho }}) > z >  - d$) - by ``2'' index, and the physical quantities, related to the dielectric solid substrate ($z <  - d$) - by ``3'' index.

Let us introduce the parameters, describing the system. In region ``1'' the system is completely described by the distribution function of charges ${f_{\bf{p}}}\left( {\bf{r}} \right)$, the potential of electric field, generated by the system of charges $\varphi _1^{(i)}\left( {\bf{r}} \right)$, the  potential of external clamping electrostatic field $\varphi _1^{(e)}\left( {\bf{r}} \right)$ and the surface profile of liquid dielectric $\xi ({\boldsymbol{\rho }})$.
In region ``2'' the system is characterized by the surface profile of liquid dielectric $\xi ({\boldsymbol{\rho }})$ and by the total electric field potential.
``Total'' means the sum of an external electric field in liquid dielectric and the field, induced by charges of region ``1''.
Region ``3'' is characterized by the total electric field potential in the solid substrate.
%\texttt{figure*}
To obtain the self-consistency equations, relating the equilibrium values of basic parameters ${f_{\bf{p}}}\left( {\bf{r}} \right)$, $\xi ({\boldsymbol{\rho }})$  and $\varphi _1^{(i)}\left( {\bf{r}} \right)$, describing the system, it is necessary to solve the problem on obtaining the maximum of the system entropy $S$

%\begin{align}\end{align}
\begin{eqnarray}
S =  - g_s\int {\frac{{d{\bf{r}}d{\bf{p}}}}{{{{\left( {2\pi \hbar } \right)}^3}}}} \left( {\bar f\ln \bar f + } \right.\left. {\left( {1 - \bar f} \right)\ln \left( {1 - \bar f} \right)} \right),
\quad
\bar f = {\left( {2\pi \hbar } \right)^3}{g_s^{ - 1}}{f_{\bf{p}}}\left( {\bf{r}} \right),
\quad
g_s = 2{S_Q} + 1
\label{1a5}
\end{eqnarray}
under the following conditions taking place.
Firstly, for a fixed external clamping field the total number of particles in the system, the total energy of the system
%\begin{widetext}\end{widetext}
\begin{eqnarray}
{E_t} = \int\limits_{{V_1}}^{} {d{\bf{r}}} \left( {\int\limits_{}^{} {d{\bf{p}}{f_{\bf{p}}}{\varepsilon _{\bf{p}}} + } Qn\left( {\frac{{\varphi _1^{(i)}}}{2} + \varphi _1^{(e)}} \right) + \frac{{{{\left( {\nabla \varphi _1^{(e)}} \right)}^2}}}{{8\pi }}} \right) + \int\limits_{{V_2}}^{} {d{\bf{r}}} \frac{{{{\left( {\nabla {\varphi _2}} \right)}^2}}}{{8\pi {\varepsilon ^{ - 1}}}} + \int\limits_{{V_3}}^{} {d{\bf{r}}} \frac{{{{\left( {\nabla {\varphi _3}} \right)}^2}}}{{8\pi \varepsilon _d^{ - 1}}} + \frac{\alpha }{2}\int {dS\left( {{{\left( {\nabla \xi } \right)}^2} + {{\left( {\kappa \xi } \right)}^2}} \right)},
\label{2a5}
\end{eqnarray}
and its total momentum ${\bf{P}} = \int {d{\bf{r}}d} {\bf{p}}{f_{\bf{p}}}\left( {\bf{r}} \right){\bf{p}}$ remain constant (Ref.~\cite{jmp2012LytvynenkoSK} gives a detailed grounding for the system energy in the form of Eq.~(\ref{1a5})).
If the system is at rest as a whole, its total momentum is zero.
Secondly, if the charges above liquid dielectric film are absent, the surface profile can not be deformed.
And thirdly, in all three regions of the described system the Poisson equation takes place.
In Eq.~(\ref{2a5}), ${V_j},\;j = 1,2,3$ denotes the volumes of regions ``1'', ``2'' and ``3'' respectively.
And the following denotations are made too: $dS = {d^2}\rho \sqrt {1 + {{\left( {{\nabla _{\boldsymbol{\rho }}}\xi \left( {\boldsymbol{\rho }} \right)} \right)}^2}}$, ${\nabla _{\boldsymbol{\rho }}} \equiv {\partial  \mathord{\left/
		{\vphantom {\partial  {\partial {\boldsymbol{\rho }}}}} \right.
		\kern-\nulldelimiterspace} {\partial {\boldsymbol{\rho }}}}$, ${\varphi _j} = \varphi _j^{(i)} + \varphi _j^{(e)}$.
The meaning of $\kappa$ value is given further on.
Besides that, the following definition of the particle density is used
\begin{equation}
n\left( {\bf{r}} \right) = \int {d{\bf{p}}} {f_{\bf{p}}}\left( {\bf{r}} \right).
\label{1.5a5}
\end{equation}
The problem on determining the conditional maximum of entropy can be reduced to the problem on unconditional minimum determining of the thermodynamical potential $\tilde \Omega$ (see Ref.~\cite{jmp2012LytvynenkoSK} for details):
%\begin{widetext}\end{widetext}
\begin{eqnarray}
\tilde \Omega  =  - S + {Y_0}E + {Y_i}{P_i} + {Y_4}N
+\int {d{\boldsymbol{\rho }}} {\lambda _\xi }\left( {\boldsymbol{\rho }} \right){\left. {\xi \left( {\boldsymbol{\rho }} \right)} \right|_{N = 0}}
+ \int {d{\bf{r}}} \lambda \left( {\bf{r}} \right)\left\{ {\Delta \varphi \left( {\bf{r}} \right) + 4\pi Qn\left( {\bf{r}} \right)} \right\},
\quad
\label{1.03a4}
\end{eqnarray}	
where ${Y_0},{Y_i},{Y_4},\lambda \left( {\bf{r}} \right),{\lambda _\xi }\left( {\boldsymbol{\rho }} \right)$ are the corresponding Lagrange multipliers to the above conditions.

The solution of such variation
problem in Ref.~\cite{jmp2012LytvynenkoSK} has the following equation
%\begin{widetext}\end{widetext}
\begin{eqnarray}
{\left( {\frac{{g_sT}}{\alpha }\int {d{\bf{p}}\frac{{ln\left( {1 - \bar f} \right)}}{{{{\left( {2\pi \hbar } \right)}^3}}}}  + \frac{\varepsilon }{{8\pi }}\left( {{{\left( {\nabla \varphi _2^{\left( e \right)}} \right)}^2} - {{\left( {\nabla {\varphi _2}} \right)}^2}} \right)} \right)_{z = \xi }} = {\kappa ^2}\xi \sqrt {1 + {{\left( {\nabla \xi } \right)}^2}}  - \nabla \left( {\frac{{\nabla \xi \left( {2 + {\kappa ^2}{\xi ^2} + 3{{\left( {\nabla \xi } \right)}^2}} \right)}}{{2\sqrt {1 + {{\left( {\nabla \xi } \right)}^2}} }}} \right),
\label{1.1a5}
\end{eqnarray}
where the distribution function of charges  ${f_{\bf{p}}}\left( {\bf{r}} \right)$ is given by
\begin{eqnarray}
{f_{\bf{p}}}\left( {\bf{r}} \right) = \theta \left( {z - \xi \left( {\boldsymbol{\rho }} \right)} \right)\frac{g_s}{{{{\left( {2\pi \hbar } \right)}^3}}}{\left\{ {1 + \exp {T^{ - 1}}\left( {{\varepsilon _{\bf{p}}} - \mu  + Q{\varphi _1}} \right)} \right\}^{ - 1}}.
\label{1.2a5}
\end{eqnarray}
$\theta \left( z \right)$ is the Heaviside step function, and   $\kappa $ function in Eq.~(\ref{1.1a5}) is defined by the expression
\begin{equation}
\kappa \left( d \right) = \sqrt {\rho {\alpha ^{ - 1}}\left( {g + f\left( d \right)} \right)},
\label{1.3a5}
\end{equation}
where $g$ is gravity acceleration, $\alpha$ is the surface tension of liquid dielectric,  $\rho$
is its density, and $f \sim {d^{ - 4}}$ is Van der Waals constant, which in the case of a massive liquid dielectric ($d \to \infty $) is negligible comparing to $g$ (see below).
In the case of a thin dielectric film the gravity force, acting on atoms of liquid dielectric becomes negligibly small comparing to Van der Waals forces (see Refs.~\cite{ufn2011Shikin,ltp2012MonarkhaS,ltp1982MonarkhaS,book1997Andrei,book2003MonarkhaK} and references therein).
E.g., such situation takes place for liquid helium films, thinner then $d \sim {10^{ - 4}}cm$~\cite{book1989ShikinM}.

Eqs.~(\ref{1.1a5}),~(\ref{1.2a5})  together with the equations for the electric field potentials, both external $\varphi _j^{(e)}$ and induced by charges $\varphi _j^{(i)}$ in all three regions of the system:
\begin{eqnarray}
\Delta \varphi _1^{(i)}({\bf{r}}) + 4\pi Q {n}\left( {\bf{r}} \right) = 0
,
\quad
\Delta \varphi _2^{(i)}({\bf{r}}) = 0,
\quad
\Delta \varphi _3^{(i)}({\bf{r}}) = 0,
\quad
\Delta \varphi _j^{(e)}({\bf{r}}) = 0,
\quad
j = 1,2,3
\label{1.4a5}
\end{eqnarray}
form a system of self-consistent equations.
Let us also note, that the first equation in Eq.~(\ref{1.4a5}), containing  $\varphi _j^{(i)}\left( {\bf{r}} \right)$ in $n\left( {\bf{r}} \right)$ through Wigner distribution function of charged fermions (see Eqs.~(\ref{1.5a5}),~(\ref{1.2a5})), is also called Thomas-Fermi equation.
The self-consistent equation system must be supplemented by the boundary conditions for the electric fields and their potentials at boundaries  $z = \xi \left( {\boldsymbol{\rho }}\right)$ and  $z =  - d$.
For the purpose of convenience it is done in the next section.

\section{Scenario of the phase transition resulting in the forming of spatially periodic structures and the boundary conditions fo the electric fields}

Scenario of the phase transition, resulting in the transformation of the surface of liquid dielectric film, is assumed as follows.
The external electric field, attracting charges to the flat surface of liquid dielectric film, causes its subsidence within the area of this field action~\cite{book1997Andrei,book2003MonarkhaK,ltp1982MonarkhaS,ufn2011Shikin,ltp2012MonarkhaS,jmp2012LytvynenkoSK}.
Moreover, the bottom of this deflection remains flat.
Therefore, the deformation of liquid dielectric surface, leaving the bottom deflection flat, can be characterized by a single parameter $\bar \xi$  (subsidence depth).
If the flat surface of undeformed dielectric is described by $z=0$ plane, the value of $\bar \xi$ should be negative,  $\bar \xi  < 0$.
Further increasing of the external electric field increases the absolute value of $\bar \xi$ and the bottom surface of the deformation remains flat up to a certain critical value of the total electric field ${E_c}$ on the dielectric surface,
\begin{equation}
{E_c} = {\left| {\frac{{\partial {\varphi _1}(z,{\boldsymbol{\rho }})}}{{\partial z}}} \right|_{z = \bar \xi }}.
\label{2.1a5}
\end{equation}
Naturally, in this case the inequality  $\left| {\bar \xi } \right| < d$ takes place, if liquid dielectric is a film, having thickness $d$ and located on a solid substrate.

Further increasing of the clamping electric field can result in the formation of a periodic structure on the surface profile of the formed deflection bottom.
Hence, the phase transition to a spatially periodic structures in this system occurs on the background of flat structure of liquid dielectric.
It should be noted, that the control parameter for this phase transition can be not only the external electric field, but also the temperature.
The described scenario can take place not only in the case of charged system, but in the case of quasi-neutral system too.
In the last case the density of charges above liquid dielectric surface is determined by the value of the external field, as mentioned above.
Thus, the density of charges above dielectric surface may be excluded from the control parameters of the phase transition (unlikely the charged system~\cite{jmp2012LytvynenkoSK}).
Consequently, in the phase transition point these two parameters (external electric field and temperature) are related by the equation, describing a certain curve.
This curve is obtained below.

According to the above scenario of the phase transition, the surface profile of liquid dielectric in a phase with lower symmetry may be represented as follows~\cite{cmp2009LytvynenkoS,jmp2012LytvynenkoSK}
\begin{equation}\label{2.2a5}
\xi ({\boldsymbol{\rho }}) = \bar \xi  + \tilde \xi ({\boldsymbol{\rho }}),
\end{equation}
where  $\tilde \xi ({\boldsymbol{\rho }})$ is the spatially inhomogeneous surface profile, formed as the result of the phase transition on the background of the flat bottom surface  $z = \bar \xi$.
Thus, the surface profile $\tilde \xi ({\boldsymbol{\rho }})$  is the order parameter of the considered phase transition.
In the symmetric phase this quantity has zero value, in the asymmetric one it describes the spatially periodic structure of the surface.
So, near the critical point from the asymmetric phase, the inequality
\begin{equation}\label{2.3a5}
\left| {\bar \xi } \right| \gg \left| {\tilde \xi ({\boldsymbol{\rho }})} \right|
\end{equation}
takes place.
Let us remind, that in the theory of phase transitions, term ``asymmetric phase'' means the phase, formed as the result of a phase transition, and this phase has lower symmetry than the initial one.
In this case the initial phase is called symmetric.
Let us also note, that in the case of Eq.~(\ref{2.3a5}) taking place in the neighborhood of the phase transition point, and the zero value of the order parameter at the point, the second order phase transition occurs~\cite{book1970Landau5}.

To describe the phase transitions, associated with the transformation of liquid dielectric surface and formation of spatially periodic structures in the researched system, we must obtain the following quantities:  $\bar \xi$, $\tilde \xi ({\boldsymbol{\rho }})$ and the distributions of charges and fields in the system as the result of the phase transition. For this purpose we use Eqs.~(\ref{1.1a5})~-~(\ref{1.4a5}),  supplemented by the boundary conditions for the characteristics of the electric field at the boundaries between the three regions.

The boundary conditions for the potentials ${\varphi _j}$
on the boundaries  $z = \xi\left( {\boldsymbol{\rho }} \right)$ and  $z =  - d$
have the form:
\begin{eqnarray}
\nonumber
{\varphi _1}\left( {\xi ,{\boldsymbol{\rho }}} \right) = {\varphi _2}\left( {\xi ,{\boldsymbol{\rho }}} \right),
\quad
{\varphi _2}\left( { - d,{\boldsymbol{\rho }}} \right) = {\varphi _3}\left( { - d,{\boldsymbol{\rho }}} \right),
\quad
{\left( {\left( {{\bf{n}}({\boldsymbol{\rho }}) \cdot \nabla } \right)\left\{ {\varepsilon {\varphi _2}(z,{\boldsymbol{\rho }}) - {\varphi _1}(z,{\boldsymbol{\rho }})} \right\}} \right)_{z = \xi }} = 0,
\\
\nonumber
\varphi _1^{\left( e \right)}\left( {\xi ,{\boldsymbol{\rho }}} \right) = \varphi _2^{\left( e \right)}\left( {\xi ,{\boldsymbol{\rho }}} \right),
\quad
\varphi _2^{\left( e \right)}\left( { - d,{\boldsymbol{\rho }}} \right) = \varphi _3^{\left( e \right)}\left( { - d,{\boldsymbol{\rho }}} \right),
\quad
%{\left( {\left( {{\bf{n}}({\boldsymbol{\rho }}) \cdot \nabla } \right)\left(\varphi _1^{\left( e \right)}(z,{\boldsymbol{\rho }})} \right)-\varepsilon {\left( {\left( {{\bf{n}}({\boldsymbol{\rho }}) \cdot \nabla } \right)\varphi _2^{\left( e \right)}(z,{\boldsymbol{\rho }})}	\right)_{z = \xi }}
{\left(
	\left( {{\bf{n}}({\boldsymbol{\rho }}) \cdot \nabla } \right)
	\left(
	\varphi _1^{\left( e \right)}\left(z,{\boldsymbol{\rho }}\right)
	-
	\varepsilon \varphi _2^{\left( e \right)}\left(z,{\boldsymbol{\rho }}\right)
	\right)\right)}
_{z = \xi }
= 0,
\\
{\left\{ {\varepsilon \frac{{\partial {\varphi _2}(z,{\boldsymbol{\rho }})}}{{\partial z}} - {\varepsilon _d}\frac{{\partial {\varphi _3}(z,{\boldsymbol{\rho }})}}{{\partial z}}} \right\}_{z =  - d}} = 0,
\quad
{\left\{ {\varepsilon \frac{{\partial \varphi _2^{\left( e \right)}(z,{\boldsymbol{\rho }})}}{{\partial z}} - {\varepsilon _d}\frac{{\partial \varphi _3^{\left( e \right)}(z,{\boldsymbol{\rho }})}}{{\partial z}}} \right\}_{z =  - d}} = 0,
\label{2.5a5}
\end{eqnarray}
where  ${\bf{n}}({\boldsymbol{\rho }})$
is the normal to the surface with profile $\xi ({\boldsymbol{\rho }})$ at  ${\boldsymbol{\rho }}$ point
\begin{equation}
{\bf{n}}({\boldsymbol{\rho }}) = \sigma \left\{ { - \frac{{\partial \xi }}{{\partial x}}, - \frac{{\partial \xi }}{{\partial y}},1} \right\},
\quad
\sigma  = {\left( {1 + {{\left( {\nabla \xi } \right)}^2}} \right)^{ - {1 \mathord{\left/
				{\vphantom {1 2}} \right.
				\kern-\nulldelimiterspace} 2}}}.
\label{2.5aa5}
\end{equation}
These boundary conditions correspond to the case, when the surface charges on the boundaries are absent.
Eq.~(\ref{2.5a5})  must be also supplemented by the limit conditions of the fields at infinity
\begin{eqnarray}
{\left| {\frac{{\partial {\varphi _1}}}{{\partial z}}} \right|_{z \to  + \infty }} <  + \infty
,
\quad
{\left| {\frac{{\partial {\varphi _3}}}{{\partial z}}} \right|_{z \to  - \infty }} <  + \infty,
\quad
{\left| {\frac{{\partial \varphi _1^{\left( e \right)}}}{{\partial z}}} \right|_{z \to  + \infty }} <  + \infty,
\quad
{\left| {\frac{{\partial \varphi _3^{\left( e \right)}}}{{\partial z}}} \right|_{z \to  - \infty }} <  + \infty.
\label{2.6a5}
\end{eqnarray}
Further on we consider the surface profile, which slightly differs from the flat one, and show the change of Eqs.~(\ref{1.1a5})~-~(\ref{1.4a5}) in this case.
In Ref.~\cite{jmp2012LytvynenkoSK} it was shown, that if the surface profile slowly changes along the coordinate, we have
\begin{equation}
\left| {\partial \xi ({\boldsymbol{\rho }})/\partial x} \right| \ll 1,
\quad
\left| {\partial \xi ({\boldsymbol{\rho }})/\partial y} \right| \ll 1.
\label{2.7a5}
\end{equation}

If Eqs.~(\ref{2.3a5}) -~(\ref{2.7a5}) take place, we can expect the distribution of charges and fields in the system to be a little different from the distributions, taking place in the case of flat dielectric surface $z = \bar \xi$.
Then, the potentials  of the external
and the total fields can be given as
\begin{eqnarray}
{\varphi _j}(z,{\boldsymbol{\rho }}) = {\bar \varphi _j}(z) + {\tilde \varphi _j}(z,{\boldsymbol{\rho }}),
\quad
\varphi _j^{(e)}(z,{\boldsymbol{\rho }}) = \bar \varphi _j^{(e)}(z) + \tilde \varphi _j^{(e)}(z,{\boldsymbol{\rho }}),
\quad
j = 1,2,3,
\label{2.9a5}
\end{eqnarray}
where  ${\bar \varphi _j}(z)$ and  $\bar \varphi _j^{(e)}(z)$ are the potentials of total and external electric fields respectively, in all three regions of the system (but not on the boundaries!) in the case of flat liquid dielectric surface  $z = \bar \xi$.
Potentials ${\tilde \varphi _j}(z,{\boldsymbol{\rho }})$ and  $\tilde \varphi _j^{(e)}(z,{\boldsymbol{\rho }})$ describe small potential perturbations in all three regions due to the surface inhomogeneity with profile $\tilde \xi ({\boldsymbol{\rho }})$.
As the potential perturbations are assumed to be week, the following inequalities take place
\begin{equation}
\left| {{{\bar \varphi }_j}(z)} \right| \gg \left| {{{\tilde \varphi }_j}(z,{\boldsymbol{\rho }})} \right|,
\quad
\left| {\bar \varphi _j^{(e)}(z)} \right| \gg \left| {\tilde \varphi _j^{(e)}(z,{\boldsymbol{\rho }})} \right|.
\label{2.10a5}
\end{equation}
Let us further assume, that the initially flat surface profile and then deformed after the phase transition  $\tilde \xi ({\boldsymbol{\rho }})$
\begin{equation}
\tilde \xi ({\boldsymbol{\rho }}) = \sum\limits_{{\bf{q}} \ne 0} {{\xi _{\bf{q}}}{e^{i{\bf{q }}{\boldsymbol{\rho }}}}},
\quad
{\xi _{\bf{q}}} = \frac{1}{{{{\left( {2\pi } \right)}^2}}}\int d {\boldsymbol{\rho }}\xi ({\boldsymbol{\rho }}){e^{ - i{\bf{q }}{\boldsymbol{\rho }}}}
\label{2.11a5}
\end{equation}
is spatially periodic.
In the case of  $\tilde \xi ({\boldsymbol{\rho }})$ periodicity (see Eqs.~(\ref{2.11a5})),  Eq.~(\ref{2.2a5}) leads to:
\begin{equation}
\bar \xi  \equiv \left\langle {\xi ({\boldsymbol{\rho }})} \right\rangle,
\quad
\tilde \xi ({\boldsymbol{\rho }}) = \xi ({\boldsymbol{\rho }}) - \left\langle {\xi ({\boldsymbol{\rho }})} \right\rangle,
\label{2.12a5}
\end{equation}
where  $\left\langle {...} \right\rangle$ is averaging over the period.

The periodic structure of $\tilde \xi ({\bf{q}})$ allows searching the potentials  ${\tilde \varphi _j}(z,{\boldsymbol{\rho }})$ (see Eq.~(\ref{2.9a5})) in the form:
\begin{eqnarray}
\nonumber
{\tilde \varphi _j}(z,{\boldsymbol{\rho }}) = \sum\limits_{{\bf{q}} \ne 0} {{{\tilde \varphi }_{j{\bf{q}}}}(z){e^{i{\bf{q }}{\boldsymbol{\rho }}}}},
\quad
{\tilde \varphi _{j{\bf{q}}}}(z) = \int\frac{d {\boldsymbol{\rho }}}{{{{\left( {2\pi } \right)}^2}}} {\tilde \varphi _j}(z,{\boldsymbol{\rho }}){e^{ - i{\bf{q }}{\boldsymbol{\rho }}}},
\\
\tilde \varphi _j^{\left( e \right)}(z,{\boldsymbol{\rho }}) = \sum\limits_{{\bf{q}} \ne 0} {\tilde \varphi _{j{\bf{q}}}^{\left( e \right)}(z){e^{i{\bf{q }}{\boldsymbol{\rho }}}}},
\quad
\tilde \varphi _{j{\bf{q}}}^{\left( e \right)}(z) = \int\frac{d {\boldsymbol{\rho }}}{{{{\left( {2\pi } \right)}^2}}} \tilde \varphi _j^{\left( e \right)}(z,{\boldsymbol{\rho }}){e^{ - i{\bf{q }}{\boldsymbol{\rho }}}}.
\label{2.13a5}
\end{eqnarray}
Taking into account  Eqs.~(\ref{2.9a5}),~(\ref{2.13a5}), we easily see that
\begin{eqnarray}
{\bar \varphi _j}(z) \equiv \left\langle {{\varphi _j}(z,{\boldsymbol{\rho }})} \right\rangle,
\quad
\left\langle {{{\tilde \varphi }_j}(z,{\boldsymbol{\rho }})} \right\rangle  = 0,
\quad
\bar \varphi _j^{(e)}(z) \equiv \left\langle {\varphi _j^{(e)}(z,{\boldsymbol{\rho }})} \right\rangle ,
\quad
\left\langle {\tilde \varphi _j^{(e)}(z,{\boldsymbol{\rho }})} \right\rangle  = 0.
\end{eqnarray}

To describe the phase transition on the scenario, described in the beginning of this section, it is necessary to determine the order parameter $\tilde \xi ({\boldsymbol{\rho }})$.
Considering the second order phase transition, we are able to obtain the order parameter  $\tilde \xi ({\boldsymbol{\rho }})$ as a function of the control parameters $T,E,{n_s}$ near the critical values  ${T_c},{E_c},{n_{sc}}$ using the perturbation theory in small parameters  $\tilde \xi ({\boldsymbol{\rho }})$, ${\tilde \varphi _j}(z,{\boldsymbol{\rho }})$  and $\tilde \varphi _j^{(e)}(z,{\boldsymbol{\rho }})$.

Taking into account Eqs.~(\ref{2.7a5})~-~(\ref{2.11a5}) after substituting Eqs.~(\ref{2.7a5}),~(\ref{2.9a5}), into Eqs.~(\ref{1.1a5})~-~(\ref{1.4a5}) and keeping the terms, linear in  $\tilde \xi ({\boldsymbol{\rho }})$,  ${\tilde \varphi _j}(z,{\boldsymbol{\rho }})$, $\tilde \varphi _j^{(e)}(z,{\boldsymbol{\rho }})$  and   $T - {T_c}$,  $E - {E_c}$,  $n - {n_{sc}}$, we obtain the equations, describing the spatial structure of the liquid dielectric surface and the distribution of charges and fields in the asymmetric phase near the critical surface (see the note above).
Let us write the equations, describing the system above the surface of liquid dielectric film  $z = \bar \xi$, i.e., in the region ``1''.
These equations are the main approximation of the described perturbation theory.
Subsequently, the charges above the liquid dielectric surface are considered to be electrons, and therefore in the corresponding formulae, we put the charge of an electron $Q =  - e$ instead of $Q$.
Due to the periodicity of small quantities $\tilde \xi ({\boldsymbol{\rho }})$,  ${\tilde \varphi _j}(z,{\boldsymbol{\rho }})$   and  $\tilde \varphi _j^{(e)}(z,{\boldsymbol{\rho }})$, see Eqs.~(\ref{2.11a5}), ~(\ref{2.13a5}), the main approximation is obtained by averaging over the period of the self-consistent equations Eqs.~(\ref{1.1a5})~-~(\ref{1.4a5}).
The components, representing the averaged values of terms, quadratic in  $\tilde \xi ({\boldsymbol{\rho }})$,  ${\tilde \varphi _j}(z,{\boldsymbol{\rho }})$   and  $\tilde \varphi _j^{(e)}(z,{\boldsymbol{\rho }})$, are small comparing to the main approximation, therefore they can be omitted.
Then, the Poisson equation in the first region in the main approximation has the form
\begin{eqnarray}
{\bar \varphi _1}^{\prime \prime }\left( z \right)= 4\pi en(z),
\quad
n(z) = \int {{d^3}p} {f_{\bf{p}}}(z),
\quad
{f_{\bf{p}}}(z) = \frac{\theta (z - \bar \xi )g_s}{{{{(2\pi \hbar )}^3}}}{\left( {1 + {e^{\frac{{{\varepsilon _{\bf{p}}} - \left( {e{{\bar \varphi }_1}(z) + \mu } \right)}}{T}}}} \right)^{ - 1}}.
\quad
\label{2.14a5}
\end{eqnarray}
The main order approximation of Eq.~(\ref{1.1a5}) in the mentioned parameters gives the equation to determine $\bar \xi$:
\begin{eqnarray}
\frac{{g_sT}}{{{{\left( {2\pi \hbar } \right)}^3}}}\int {d{\bf{p}}} {\left. {\ln \left( {1 - \frac{{{{\left( {2\pi \hbar } \right)}^3}}}{g}{f_{\bf{p}}}(z)} \right)} \right|_{z = \bar \xi }}
= \frac{\varepsilon }{{8\pi }}{\left( {{{\left( {{{\bar \varphi }_2}^\prime \left( z \right)} \right)}^2} - {{\left( {\bar \varphi {{_2^{\left( e \right)}}^\prime }\left( z \right)} \right)}^2}} \right)_{z = \bar \xi }} + \alpha {\kappa ^2}\bar \xi.
\label{2.15a5}
\end{eqnarray}
Due to the absence of charges in regions ``2'' and ``3'', the equations for ${\bar \varphi _2}(z)$ and ${\bar \varphi _3}(z)$ potentials have the form:
\begin{equation}
{\bar \varphi _2}^{\prime \prime }\left( z \right) = 0,
\quad
{\bar \varphi _3}^{\prime \prime }\left( z \right) = 0.
\label{2.16a5}
\end{equation}
Let us remind, that the external field potentials $\bar \varphi _j^{\left( e \right)}(z)$  in all three regions are described by the same equations (Laplace equations):
\begin{equation}
\bar \varphi {_j^{\left( e \right)\prime \prime}}\left( z \right) = 0,
\quad
j=1,2,3.
\label{2.17a5}
\end{equation}
To make the system of Eqs.~(\ref{2.14a5})~-~(\ref{2.17a5}) self-contained, the same averaging procedure is applied to the boundary conditions Eq.~(\ref{2.5a5}).
As the result, we obtain the relation between the total potentials  ${\bar \varphi _j}(z)$ and the potentials of external electric field  $\bar \varphi _j^{(e)}(z)$ on the boundaries of three regions:
\begin{eqnarray}
\nonumber
{\bar \varphi _1}\left( {\bar \xi } \right) = {\bar \varphi _2}\left( {\bar \xi } \right),
\quad
{\bar \varphi '_1}\left( {z = \bar \xi } \right) = \varepsilon {\bar \varphi '_2}\left( {z = \bar \xi } \right),
\quad
{\bar \varphi _2}\left( { - d} \right) = {\bar \varphi _3}\left( { - d} \right),
\quad
{\left( {\varepsilon {{\bar \varphi '}_2} - {\varepsilon _d}{{\bar \varphi '}_3}} \right)_{z =  - d}} = 0,
\\
\bar \varphi _1^{\left( e \right)}\left( {\bar \xi } \right) = \bar \varphi _2^{\left( e \right)}\left( {\bar \xi } \right),
\quad
{\left( {{\varepsilon _d}\bar \varphi {{_3^{\left( e \right)}}^\prime } - \varepsilon \bar \varphi {{_2^{\left( e \right)}}^\prime }} \right)_{z =  - d}} = 0,
\quad
\bar \varphi _2^{\left( e \right)}\left( { - d} \right) = \bar \varphi _3^{\left( e \right)}\left( { - d} \right),
\quad
{\left( {\bar \varphi {{_1^{\left( e \right)}}^\prime } - \varepsilon \bar \varphi {{_2^{\left( e \right)}}^\prime }} \right)_{z = \bar \xi }} = 0.
\label{2.18a5}
\end{eqnarray}

It is easy to see, that the solution of the problem on the phase transition description starts from the solving the equations of the main approximation Eqs.~(\ref{2.14a5})~-~(\ref{2.17a5}) with the boundary conditions Eq.~(\ref{2.18a5}).
This procedure provides obtaining the distribution of charges and fields in the system in the case of flat surface of liquid dielectric, which is given by the equation $z = \bar \xi$. The value of  $\bar \xi$, obtained from Eq.~(\ref{2.15a5}), determines the subsidence level of the flat dielectric surface due to the pressure of charges (electrons) on it.

To obtain the critical parameters of the considered phase transition, the higher orders of perturbation theory must be involved.
Below, we formulate the system of self-consistent Eqs.~(\ref{1.1a5})~-~(\ref{1.4a5}) in the first order of this theory.
To simplify the further calculations, we assume the resulting periodic structure to be one-dimensional with the period along the $x$  axis equal to   $a$, so  $q = {q_x} = \frac{{2\pi }}{a}$.
So, further on, in place of vector  ${\bf{q}}$, directed along  $x$ axis, we write its corresponding projection $q$.
Let us consider the quantities ${\tilde \xi _q}$, ${\tilde \varphi _{jq}}(z)$  and $\tilde \varphi _{jq}^{\left( e \right)}(z)$ to have the following form:
\begin{eqnarray}
{\tilde \xi _q}(z) = \sum\limits_{l = 1}^\infty  {\tilde \xi _q^{\left( l \right)}},
\quad
{\tilde \varphi _{jq}}(z) = \sum\limits_{l = 1}^\infty  {\tilde \varphi _{jq}^{\left( l \right)}(z)},
\quad
\tilde \varphi _{jq}^{\left( e \right)}(z) = \sum\limits_{l = 1}^\infty  {\tilde \varphi _{jq}^{\left( e \right)\left( l \right)}(z)},
\label{2.19a5}
\end{eqnarray}
where
\begin{eqnarray}
\tilde \xi _q^{\left( 1 \right)} = \tilde \xi _{{q_0}}^{\left( 1 \right)}\left( {\Delta \left( {q - {q_0}} \right) + \Delta \left( {q + {q_0}} \right)} \right),
\quad
\tilde \xi _q^{\left( 2 \right)} = \tilde \xi _{2{q_0}}^{\left( 2 \right)}\left( {\Delta \left( {q - 2{q_0}} \right) + \Delta \left( {q + 2{q_0}} \right)} \right),
\label{2.20a5}
\end{eqnarray}
and the values of $\tilde \varphi _{j{\bf{q}}}^{\left( 1 \right)}\left( z \right)$ and $\tilde \varphi _{j{q_0}}^{\left( 1 \right)}\left( z \right)$, $\tilde \varphi _{j{\bf{q}}}^{\left( 2 \right)}\left( z \right)$ and $\tilde \varphi _{j2{q_0}}^{\left( 2 \right)}\left( z \right)$, $\tilde \varphi _{j{\bf{q}}}^{\left( e \right)\left( 1 \right)}\left( z \right)$ and $\tilde \varphi _{j{q_0}}^{\left( e \right)\left( 1 \right)}\left( z \right)$, $\tilde \varphi _{j{\bf{q}}}^{\left( e \right)\left( 2 \right)}\left( z \right)$ and $\tilde \varphi _{j2{q_0}}^{\left( e \right)\left( 2 \right)}\left( z \right)$ are related in similar to Eq.~(\ref{2.20a5}) way.
In these formulae $\Delta \left( q \right)$ is the Kronecker symbol
\begin{eqnarray}
\nonumber
\Delta \left( q \right) = \left\{ \begin{array}{l}
0,\quad \,q \ne 0 \\
1,\,\quad q = 0 \\
\end{array} \right..
\end{eqnarray}
In Eq.~(\ref{2.19a5}) we assume the appeared periodic structure to be one-dimensional with a period along $x$ axis equal to $a$, so $q = {q_x} = {{2\pi } \mathord{\left/
		{\vphantom {{2\pi } a}} \right.
		\kern-\nulldelimiterspace} a}.$

We also assume, that  $\tilde \varphi _{jq}^{}(z) = \tilde \varphi _{j - q}^{}(z)$ and  $\tilde \xi _q^{} = \tilde \xi _{ - q}^{}$, thereby considering the real values of these quantities, so
\begin{eqnarray}
\nonumber
\tilde \xi (x) = 2\sum\limits_{l = 1}^{ + \infty } {\tilde \xi _{}^{\left( l \right)}\cos l{q_0}x},
\quad
\tilde \varphi _j^{}\left( {x,z} \right) = 2\sum\limits_{l = 1}^{ + \infty } {\tilde \varphi _j^{\left( l \right)}(z)\cos l{q_0}x},
\quad
\tilde \varphi _j^{\left( e \right)}\left( {x,z} \right) = 2\sum\limits_{l = 1}^{ + \infty } {\tilde \varphi _j^{\left( e \right)\left( l \right)}(z)\cos l{q_0}x}.
\end{eqnarray}
Then, the linear approximation for Eqs.~(\ref{1.1a5}),~(\ref{1.4a5}),~(\ref{2.5a5}) in small values of the first harmonics of ${\tilde \xi _q}$ and ${\tilde \varphi _{jq}}(z)$ has the following form:
%\begin{widetext}
	\begin{eqnarray}
	\nonumber
	\frac{{{\partial ^2}{{\tilde \varphi }^{\left( 1 \right)}}_1}}{{\partial {z^2}}} - q_0^2{\tilde \varphi ^{\left( 1 \right)}}_1 = 4\pi {e^2}\frac{{\partial n}}{{\partial \mu }}{\tilde \varphi ^{\left( 1 \right)}}_1
	,
	\quad
	\frac{{{\partial ^2}{{\tilde \varphi }^{\left( 1 \right)}}_2}}{{\partial {z^2}}} - q_0^2{\tilde \varphi ^{\left( 1 \right)}}_2 = 0
	,
	\quad
	\frac{{{\partial ^2}{{\tilde \varphi }^{\left( 1 \right)}}_3}}{{\partial {z^2}}} - q_0^2{\tilde \varphi ^{\left( 1 \right)}}_3 = 0,
	\quad
	{\left( {{{\tilde \varphi }^{\left( 1 \right)}}_2 - {{\tilde \varphi }^{\left( 1 \right)}}_3} \right)_{z =  - d}} = 0,
	\\
	\nonumber
	{\left( {\left( {{{\bar \varphi '}_1} - {{\bar \varphi '}_2}} \right){{\tilde \xi }^{\left( 1 \right)}} + {{\tilde \varphi }_1}^{\left( 1 \right)} - {{\tilde \varphi }_2}^{\left( 1 \right)}} \right)_{z = \bar \xi }} = 0,
	\;
	{\left( {{{\bar \varphi ''}_1}{{\tilde \xi }^{\left( 1 \right)}} + \frac{{\partial {{\tilde \varphi }_1}^{\left( 1 \right)}}}{{\partial z}} - \varepsilon \frac{{\partial \tilde \varphi _2^{\left( 1 \right)}}}{{\partial z}}} \right)_{z = \bar \xi }} = 0,
	\;
	{\left( {\varepsilon \frac{{\partial {{\tilde \varphi }^{\left( 1 \right)}}_2}}{{\partial z}} - {\varepsilon _d}\frac{{\partial {{\tilde \varphi }^{\left( 1 \right)}}_3}}{{\partial z}}} \right)_{z =  - d}} = 0.
	\\
	{\left( {en\left( {{{\tilde \varphi }^{\left( 1 \right)}}_1 + {{\bar \varphi '}_1}{{\tilde \xi }^{\left( 1 \right)}}} \right)} \right)_{z = \bar \xi }} + \frac{\varepsilon }{{4\pi }}{\left( {{{\bar \varphi '}_2}\frac{{\partial {{\tilde \varphi }^{\left( 1 \right)}}_2}}{{\partial z}} - \bar \varphi {{_2^{\left( e \right)}}^\prime }\frac{{\partial \tilde \varphi _2^{\left( e \right)\left( 1 \right)}}}{{\partial z}}} \right)_{z = \bar \xi }} + \alpha {\tilde \xi ^{\left( 1 \right)}}\left( {{\kappa ^2} + q_0^2\left( {1 + \frac{{{\kappa ^2}{{\bar \xi }^2}}}{2}} \right)} \right) = 0.
	\quad
	\label{2.21a5}
	\end{eqnarray}

	Similarly, the first approximation of the considered perturbation theory for Eq.~(\ref{1.1a5}) for the external potential, supplemented by the appropriate boundary conditions, has the form
	\begin{eqnarray}
	\nonumber
	\frac{{{\partial ^2}{{\tilde \varphi }^{\left( e \right)\left( 1 \right)}}_j}}{{\partial {z^2}}} - q_0^2\tilde \varphi _j^{\left( e \right)\left( 1 \right)} = 0,
	\quad
	j = 1,2,3,
	\quad
	{\left( {\left( {\bar \varphi {{_1^{\left( e \right)}}^\prime } - \bar \varphi {{_2^{\left( e \right)}}^\prime }} \right){{\tilde \xi }^{\left( 1 \right)}} + {{\tilde \varphi }_1}^{\left( e \right)\left( 1 \right)} - {{\tilde \varphi }_2}^{\left( e \right)\left( 1 \right)}} \right)_{z = \bar \xi }} = 0,
	\\
	{\left( {\tilde \varphi _2^{\left( e \right)\left( 1 \right)} - \tilde \varphi _3^{\left( e \right)\left( 1 \right)}} \right)_{z =  - d}} = 0,
	\quad
	{\left( {\frac{{\partial {{\tilde \varphi }_1}^{\left( e \right)\left( 1 \right)}}}{{\partial z}} - \varepsilon \frac{{\partial \tilde \varphi _2^{\left( e \right)\left( 1 \right)}}}{{\partial z}}} \right)_{z = \bar \xi }} = 0,
	\quad
	{\left( {\varepsilon \frac{{\partial \tilde \varphi _2^{\left( e \right)\left( 1 \right)}}}{{\partial z}} - {\varepsilon _d}\frac{{\partial \tilde \varphi _3^{\left( e \right)\left( 1 \right)}}}{{\partial z}}} \right)_{z =  - d}} = 0.
	\label{2.23a5}
	\end{eqnarray}
%\end{widetext}

Eqs.~(\ref{2.21a5})~-~(\ref{2.23a5}) allow solving the above problem on describing the phase transition, associated with the formation of spatially periodic structures above the liquid dielectric film surface in the studied system.
We also note, that the values of $T$  and  $E$, included in the coefficients, multiplied by small deviations ${\tilde \xi ^{\left( 1 \right)}},{\tilde \varphi _j}^{\left( e \right)\left( 1 \right)},{\tilde \varphi _j}^{\left( 1 \right)}$,  are related by the already mentioned critical surface curve.
The linear approximation of the discussed equations does not contain the terms, proportional to $T - {T_c}$  and $E - {E_c}$,  because they have a higher order of smallness. This fact causes obtaining the higher orders of the perturbation theory to calculate the dependence of the order parameter ${\tilde \xi ^{\left( 1 \right)}}$  on the control parameters  $T,E$ near the critical surface.
The following sections are devoted to solving this problem, and to solving the system of Eqs.~(\ref{2.14a5})~-~(\ref{2.18a5}) and Eqs.~(\ref{2.21a5})~-~(\ref{2.23a5}).

In this paper the system of charges is not considered to be localized in any plane, unlikely in Refs.~\cite{prl1979GrimesA,ltp1982MonarkhaS,ltp2012MonarkhaS,book2003MonarkhaK}.
These papers are concerned with the 2D hexagonal crystal structures, formed by the electrons above liquid helium surface.
The exceptional cases are those, where the so-called ``dimple'' crystals are described.
As seen from the above problem, this work describes a 3D system of charges by the distribution function, depending on the coordinates of the half-space above the liquid dielectric surface.
Further on, it is shown, that  present work considers the spatially periodic structure along the directions, parallel to the plane $\left( {x,y} \right)$, and these structures are caused by the spatial periodicity of the surface profile of the liquid dielectric film.

\section{Distribution of charges and field in electro-neutral system above flat surface of liquid dielectric}

The solution of Eqs.~(\ref{2.14a5})~-~(\ref{2.17a5}) is obtained in terms of the methods, proposed in Ref.~\cite{jmp2012LytvynenkoSK} (see also Ref.~\cite{cmp2009LytvynenkoS,jps2015SlyusarenkoL}), where the similar problem was considered in the case of a non-degenerate gas of charges above a flat solid dielectric boundary.
The difference of system description in the cases of solid and liquid flat dielectric boundaries, is that the $z$ coordinate of solid flat dielectric surface stays fixed, and the liquid dielectric surface ``sinks'' under the influence of the additional pressure, created by charges (see below).
A similar problem was solved in Ref.~\cite{jetp2000DyugaevGO}, where the authors obtained the distribution of non-degenerate electron gas inside a flat capacitor with plates, covered by a flat dielectric layer.
However, in present paper the case of general statistics of Fermi-particles Eq.~(\ref{1.2a5}) is considered.
To solve the first equation in Eq.~(\ref{2.14a5}) it is convenient to rewrite it in the following form:
\begin{equation}
{\bar \varphi _1}^{\prime \prime }\left( z \right) = 4\pi e\nu \int\limits_0^\infty  {d\varepsilon {\varepsilon ^{1/2}}{{\left( {1 + \exp \frac{{\varepsilon  - \psi }}{T}} \right)}^{ - 1}}},
\label{3.1a5}
\end{equation}
where the following notations are introduced
\begin{eqnarray}
\psi \left( z \right) \equiv \mu  + e{\bar \varphi _1}\left( z \right),
\quad
\nu  = \sqrt 2 {\pi ^{ - 2}}a_0^{ - 3/2}{e^{ - 3}}.
\label{3.2a5}
\end{eqnarray}
We also take into account that $S_Q=1/2$ and ${a_0} \equiv {{{\hbar ^2}} \mathord{\left/
		{\vphantom {{{\hbar ^2}} {\left( {m{e^2}} \right)}}} \right.
		\kern-\nulldelimiterspace} {\left( {m{e^2}} \right)}}$ is first Bohr radius.
$\psi$ is usually called as an electrochemical potential.

\begin{figure}
	\begin{minipage}{.5\textwidth}
		\includegraphics[scale=0.70]{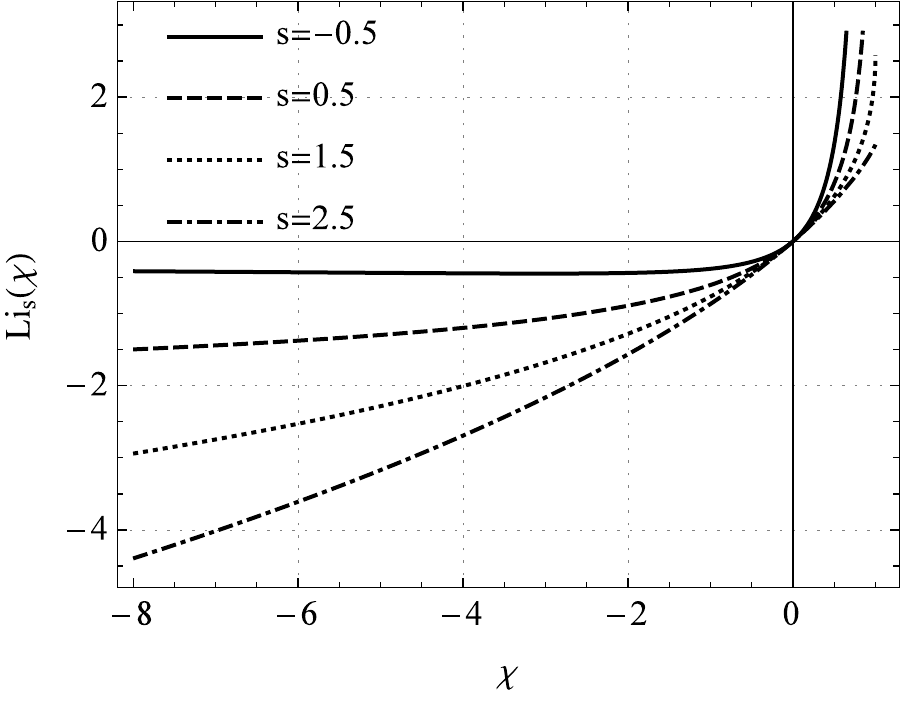}
		\caption{\label{fig:pic2a5}  Plots of  $L{i_s}\left( \chi  \right)$ for different values of $s$.}
	\end{minipage}
	\hfill
	\begin{minipage}{.5\textwidth}
		\includegraphics[scale=0.9]{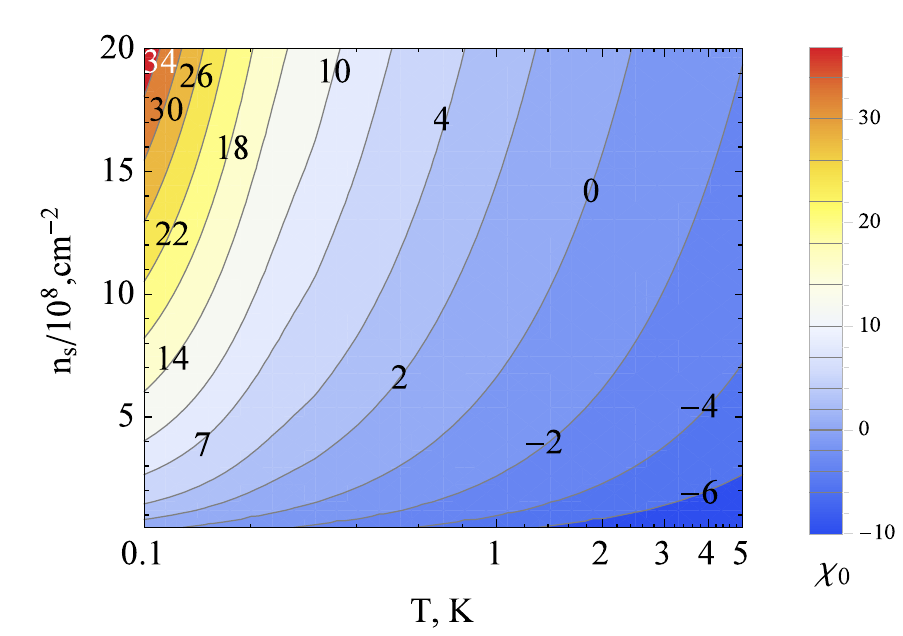}
		\caption{\label{fig:pic3a5}  The dependence of ${\chi _0} = \chi \left( {z = \bar \xi } \right)$ on $n_s$ and $T$.}
	\end{minipage}
\end{figure}

The order of Eq.~(\ref{3.1a5}) can be lowered (see, e.g., Ref.~\cite{jmp2012LytvynenkoSK})
\begin{equation}
{\bar \varphi _1}^\prime \left( z \right) =  - {\left\{ {\frac{{16\pi }}{3}\nu \int\limits_0^\infty  {\frac{{d\varepsilon {\varepsilon ^{3/2}}}}{{1 + {e^{\frac{{\varepsilon  - \psi }}{T}}}}}}  + {C_1}} \right\}^{1/2}},
\label{3.3a5}
\end{equation}
where $C_1$ is an arbitrary integration constant. The sign before the square root in this equation is chosen, assuming the force, acting on a negative charge at $z > \bar \xi$, pressing it to the dielectric surface.
Eq.~(\ref{3.3a5}) is simplified in the case of quasi-neutral system.
Indeed, the quasi-neutrality condition assumes the absence of particles at infinity (see Eq.~(\ref{2.14a5}))
\begin{equation}
\nonumber
\mathop {\lim }\limits_{z \to  + \infty } {\left( {\exp \beta \left( {\varepsilon  - \psi } \right) + 1} \right)^{ - 1}} = 0.
\end{equation}
Taking this fact into account, Eq.~(\ref{3.3a5}) turns to the following form
\begin{equation}
\nonumber
\mathop {\lim }\limits_{z \to  + \infty } {\bar \varphi _1}^\prime \left( z \right) =  - \sqrt {{C_1}}.
\end{equation}
In the case of quasi-neutrality of the system the total electric field vanishes at infinity, so $C_1=0$.
Taking into account Eqs.~(\ref{3.2a5}),~(\ref{3.3a5}), we have
\begin{equation}
\frac{{\partial \chi }}{{\partial z}} =  - \frac{{{2^{{5 \mathord{\left/
						{\vphantom {5 4}} \right.
						\kern-\nulldelimiterspace} 4}}}}}{{{a_0}}}{\left( {\frac{{T{a_0}}}{{\pi {e^2}}}} \right)^{{1 \mathord{\left/
				{\vphantom {1 4}} \right.
				\kern-\nulldelimiterspace} 4}}}{\left( { - L{i_{{5 \mathord{\left/
						{\vphantom {5 2}} \right.
						\kern-\nulldelimiterspace} 2}}}\left( { - {e^\chi }} \right)} \right)^{{1 \mathord{\left/
				{\vphantom {1 2}} \right.
				\kern-\nulldelimiterspace} 2}}},
\quad
\chi  = {\psi  \mathord{\left/
		{\vphantom {\psi  T}} \right.
		\kern-\nulldelimiterspace} T}.
\label{3.4a5}
\end{equation}
Here we use a special polylogarithmic function   $L{i_s}\left( \chi  \right)$, related to the Fermi-Dirac integral
${I_s}\left( \chi  \right)$ by the following expression
\begin{eqnarray}
{I_s}\left( \chi  \right) =  - L{i_{s + 1}}\left( { - {e^\chi }} \right),
\quad
{I_s}\left( \chi  \right) = \frac{1}{{\Gamma \left( {s + 1} \right)}}\int\limits_0^\infty  {\frac{{{x^s}dx}}{{1 + {e^{x - \chi }}}}}.
\label{3.5a5}
\end{eqnarray}
The plots of polylagorithmic functions of different orders $s$ are presented on Fig.~\ref{fig:pic2a5}
Integration of Eq.~(\ref{3.4a5}) requires applying numerical methods.
In certain cases they are used further on. However, a set of important results, such as the relation between the critical parameters of the phase transition, the quasi-neutrality condition, etc., can be expressed analytically without the calculation of explicit dependence on $z$ of contained quantities.
Nevertheless, the calculus of ${E_1}\left( z \right)$, $n\left( z \right)$ and $\chi \left( z \right)$ is made below.

As shown in Ref.~\cite{cmp2009LytvynenkoS}, in general case the gas of charges can be degenerate near the dielectric surface and non-degenerate at large distance from it.
The condition of gas degeneracy is defined by the system temperature, particle density and the external clamping field value.
Unlike Refs.~\cite{jmp2012LytvynenkoSK, jps2015SlyusarenkoL}, the present paper considers the general case of the distribution function Eqs.~(\ref{1.2a5}),~(\ref{2.14a5}).

According to Eqs.~(\ref{2.14a5}),~(\ref{3.5a5}), the expression for density, as a function of $\chi$, has the following form (see Refs.~\cite{jmp2012LytvynenkoSK,cmp2009LytvynenkoS}):
\begin{equation}
n =  - {\left( {\frac{{T{a_0}}}{{\pi {e^2}}}} \right)^{{3 \mathord{\left/
				{\vphantom {3 2}} \right.
				\kern-\nulldelimiterspace} 2}}}\frac{{L{i_{{3 \mathord{\left/
						{\vphantom {3 2}} \right.
						\kern-\nulldelimiterspace} 2}}}\left( { - {e^\chi }} \right)}}{{\sqrt 2 a_0^3}}.
\label{3.6a5}
\end{equation}
Eq.~(\ref{3.6a5}) is the result of normalization of the distribution function Eq.~(\ref{2.14a5}) on the total number of charges $N$ above the surface $S$:
\begin{eqnarray}
\int {d{\bf{x}}} \sum\limits_{\bf{p}} {{f_{\bf{p}}}({\bf{x}})}  = \int {d{\boldsymbol{\rho }}} \int\limits_0^\infty  {dz\int {\frac{{d{\bf{p}}{f_{\bf{p}}}(z)}}{{{{\left( {2\pi \hbar } \right)}^3}}}} }
=S\int\limits_0^\infty  {dzn\left( z \right)}  = N.
\label{*a5}
\end{eqnarray}
In fact, Eq.~(\ref{*a5}) is approximate, because a certain part of charges belongs to the spatially periodic structure of liquid dielectric surface.
However, according to Eqs.~(\ref{2.3a5}),~(\ref{2.10a5}), the number of such charges is small comparing to the total number of charges $N$.

Let us establish the relation between the value of chemical potential $\mu$, contained in $\psi \left( z \right)$ (see Eq.~(\ref{3.2a5})), and the number of charges, located above the unit of flat liquid dielectric surface area $n_s$:
\begin{eqnarray}
{n_s} = \int\limits_\xi ^\infty  {dzn(z)}  = \frac{N}{S}.
\label{3.7a5}
\end{eqnarray}
Taking into account Eqs.~(\ref{3.5a5}),~(\ref{3.6a5}) we change in Eq.~(\ref{3.7a5}) the integration variable from $z$ to $\chi$ and obtain
\begin{eqnarray}
{n_s} = {\left( {\frac{{T{a_0}}}{{\pi {e^2}}}} \right)^{{5 \mathord{\left/
				{\vphantom {5 4}} \right.
				\kern-\nulldelimiterspace} 4}}}\frac{{{{\left( { - L{i_{{5 \mathord{\left/
									{\vphantom {5 2}} \right.
									\kern-\nulldelimiterspace} 2}}}\left( { - {e^{{\chi _0}}}} \right)} \right)}^{{1 \mathord{\left/
						{\vphantom {1 2}} \right.
						\kern-\nulldelimiterspace} 2}}}}}{{{2^{{3 \mathord{\left/
						{\vphantom {3 4}} \right.
						\kern-\nulldelimiterspace} 4}}}a_0^2}}.
\label{3.8a5}
\end{eqnarray}
Let us emphasize, that Eq.~(\ref{3.8a5}) is obtained, using the following assumption
\begin{eqnarray}
\mathop {\lim }\limits_{z \to  + \infty } \chi  =  - \infty,
\label{3.8aa5}
\end{eqnarray}
meaning the absence of charges at infinity. Otherwise, the value of ${\left( { - L{i_{{5 \mathord{\left/
						{\vphantom {5 2}} \right.
						\kern-\nulldelimiterspace} 2}}}\left( { - {e^{\chi \left( \infty  \right)}}} \right)} \right)^{{1 \mathord{\left/
				{\vphantom {1 2}} \right.
				\kern-\nulldelimiterspace} 2}}}$
and, hence, the distribution charges at infinity is not equal to zero.
This fact contradicts the made above assumption on the absence of charges at infinity and the quasi-neutrality condition.

Eq.~(\ref{3.8a5}) provides the ability of numeric calculation of the dependence of $\chi \left( {z = \bar \xi } \right) = {\chi _0}$
on the temperature $T$ and the number of charges above dielectric surface area unit $n_s$.
This dependence is presented on Fig.~\ref{fig:pic3a5}.

Noticing, that ${E_1}\left( z \right) =  - {\bar \varphi _1}^\prime \left( z \right)$ and taking into account Eq.~(\ref{3.4a5}), we obtain
\begin{eqnarray}
4\pi e{n_s} = {E_1}\left( {z = \bar \xi } \right) \equiv {E_0}.
\label{3.9a5}
\end{eqnarray}
Eq.~(\ref{3.9a5}) shows, that the value of electric field on the dielectric surface is equivalent to the value of  electric field inside the plane capacitor with oppositely charged plates with a surface charge density equal in absolute value to $\sigma  = e{n_s}$.

\begin{figure}
	\begin{minipage}{.49\textwidth}
		\includegraphics[scale=0.82]{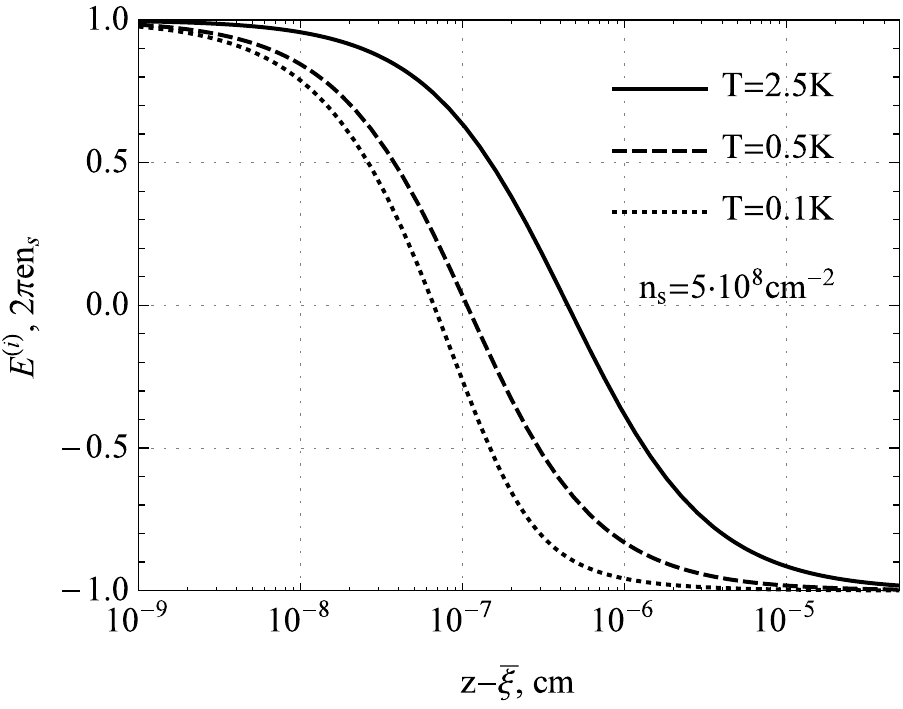}
		\caption{\label{fig:pic3aa5} Electric field ${E^{\left( i \right)}}\left( z \right)$, generated by charges for three pairs of $T$ and $n_s$ values.}
	\end{minipage}
	\hfill
	\begin{minipage}{.49\textwidth}
		\centering
		\includegraphics[scale=0.90]{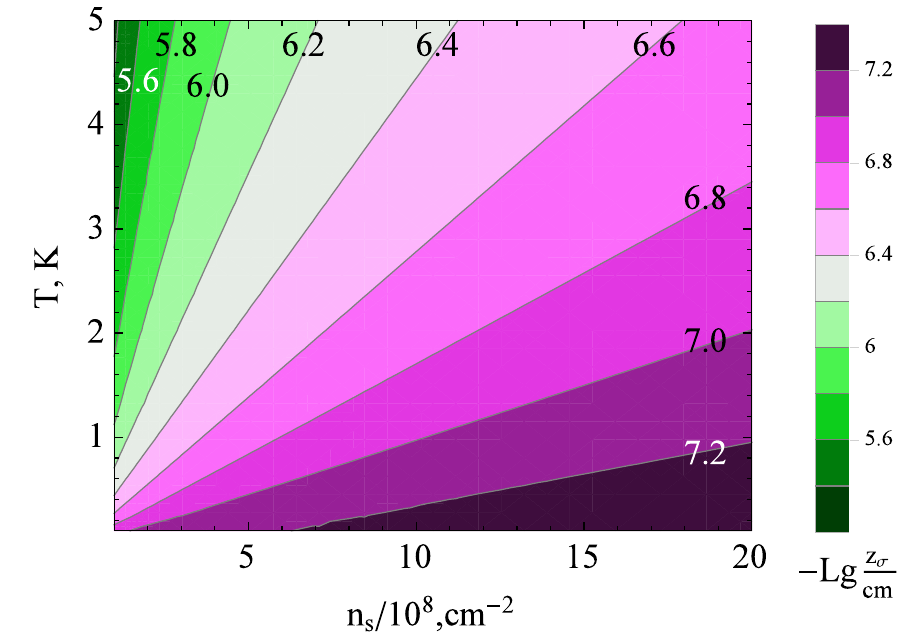}
		\caption{\label{fig:pic3ba5} $z_\sigma$ distance from the dielectric surface, where the electric field, induced by charges, vanishes.}
	\end{minipage}
\end{figure}

To solve the equation system Eqs.~(\ref{2.14a5}) -~(\ref{2.17a5}) using the boundary conditions Eq.~(\ref{2.18a5}), we have to obtain the relation between the values of the external electric field $E_1^{\left( e \right)}\left( z \right) =  - \bar \varphi {_1^{\left( e \right)\prime }}\left( z \right)$ and the total electric field ${E_1}\left( z \right) =  - {\bar \varphi _1}^\prime \left( z \right)$.
If charges are present above the dielectric surface, they make a contribution to the total  electric field $E_1^{}\left( z \right) = E_1^{\left( i \right)}\left( z \right) + E_1^{\left( e \right)}\left( z \right)$ by inducing their own field $E_1^{\left( i \right)}\left( z \right) =  - \bar \varphi {_1^{\left( i \right)\prime}}
\left( z \right)$.
Using the density distribution of charges Eq.~(\ref{3.6a5}), we can easily calculate this field value.
The z-component of electric field at the point, having $z$ coordinate, which is produced by the elementary volume of charges $dx'dy'dz'$, located at the point $\left( {x',y',z'} \right)$, has the form:
\begin{eqnarray}
dE_{1z}^{\left( i \right)}\left( z \right) =  - \frac{{\left( {z - z'} \right)en\left( {z'} \right)dx'dy'dz'}}{{{{\left( {{{x'}^2} + {{y'}^2} + {{\left( {z - z'} \right)}^2}} \right)}^{{3 \mathord{\left/
						{\vphantom {3 2}} \right.
						\kern-\nulldelimiterspace} 2}}}}}.
\label{3.10a5}
\end{eqnarray}
The minus sign before the ratio  in Eq.~(\ref{3.10a5}) shows that the particles are negatively charged.
Assuming the system infinity along $\left( {x,y} \right)$ coordinates, the integration of the corresponding components of electric field $dE_{1x}^{\left( i \right)},dE_{1y}^{\left( i \right)}$ in the total $V_1$ volume  gives zero.
That is why, the electric field generated by charges has only z-component. It can be obtained by applying the integration procedure $E_{1z}^{\left( i \right)}\left( z \right) = \int\limits_{{V_1}} {dE_{1z}^{\left( i \right)}\left( z \right)}$, which results in the following expression:
\begin{eqnarray}
E_{1z}^{\left( i \right)}\left( z \right) =  - 2\pi e\left( {\int\limits_{\bar \xi }^z {dz'n\left( {z'} \right) - } \int\limits_z^\infty  {dz'n\left( {z'} \right)} } \right).
\quad
\label{3.120a5}
\end{eqnarray}
Applying numeric integration methods to Eq.~(\ref{3.120a5}), we obtain Fig.~\ref{fig:pic3aa5}, which illustrates the ${E^{\left( i \right)}}\left( z \right)$ dependence for three pairs of $T$ and $n_s$ values.
The reason for choosing these values is explained below in the section.
Let us give a simple physical interpretation of Eq.~(\ref{3.120a5}) by considering its limit cases.
According to Eqs.~(\ref{3.7a5}),~(\ref{3.120a5}), we have $\mathop {\lim }\limits_{z \to \bar \xi } E_{1z}^{\left( i \right)}\left( z \right) = 2\pi e{n_s}$ and $\mathop {\lim }\limits_{z \to  + \infty } E_{1z}^{\left( i \right)}\left( z \right) =  - 2\pi e{n_s}$.
These values of electric field, produced by the system of charges, are equivalent to the similar values of infinite plate, having a surface charge density $\sigma  = e{n_s}$.
According to Eqs.~(\ref{1.4a5}), the value of external electric field in its main approximation is constant, $E_1^{\left( e \right)} = {E_1}\left( z \right) - E_1^{\left( i \right)}\left( z \right) \equiv E$.
The values of external and total electric fields coincide in a certain point ${z_\sigma }$, where $E = {E_1}\left( {{z_\sigma }} \right)$ and $E_1^{\left( i \right)}\left( {{z_\sigma }} \right) = 0$.
${z_\sigma }$ is evaluated numerically and it is presented on Fig.~\ref{fig:pic3ba5}.
At fixed value of ${n_s} = 5 \cdot {10^8}c{m^{ - 2}}$ we obtain ${z_\sigma } \approx 4,47 \cdot {10^{ - 7}}cm$ for $T = 2,5K$, ${z_\sigma } \approx 1,07 \cdot {10^{ - 7}}cm$ for $T = 0,5K$ and ${z_\sigma } \approx 6,66 \cdot {10^{ - 8}}cm$ for $T = 0,1K$.
However, rather than evaluating ${z_\sigma }$ to obtain $E$, it is convenient to calculate the following limit:
\begin{eqnarray}
E = \mathop {\lim }\limits_{z \to  + \infty } \left( {{E_1}\left( z \right) - E_1^{\left( i \right)}\left( z \right)} \right) = {E_\infty } + 2\pi e{n_s},
\quad
\label{3.12aa5}
\end{eqnarray}
where $\mathop {\lim }\limits_{z \to  + \infty } {E_1}\left( z \right) = {E_\infty }$.
In the considered electro-neutral case, we have ${E_\infty } = 0$.
Consequently, Eqs.~(\ref{3.12aa5}) turns to the following form:
\begin{eqnarray}
E = 2\pi e{n_s}.
\label{3.12ba5}
\end{eqnarray}

Let us make an important remark.
The obtained values of electric fields Eqs.~(\ref{3.9a5}),~(\ref{3.12ba5}) are used in the present paper approach.
But comparing them to the experimental data of some papers (e.g., see Refs.~\cite{book2003MonarkhaK,ltp1982MonarkhaS,
	ltp2012MonarkhaS,ufn2011Shikin}) can lead to some quantitative mismatch.
This is caused by the experimental measurement of the potential difference between the plates of the capacitor, creating the external field, but not the measurement of this field value.
Let us obtain the relation between the potential difference of the capacitor plates and $n_s$ parameter in the quasi-neutral case.
In our system the solid substrate can be considered as a lower capacitor plate, dived into liquid dielectric on $d$ depth.
As far as the upper capacitor plate is concerned, in our system it is located at the infinite distance from the dielectric surface.
If the charges between the capacitor plates are absent, the applied voltage has the form $\Delta {U^{\left( e \right)}} = \bar \varphi _1^{\left( e \right)}\left( { + \infty } \right) - \bar \varphi _2^{\left( e \right)}\left( { - d} \right)$.
This potential difference produces the electric field $E$ above liquid dielectric and ${{E} \mathord{\left/
		{\vphantom {{E} \varepsilon }} \right.
		\kern-\nulldelimiterspace} \varepsilon }$ below it.
Then the charges are injected into the volume above liquid dielectric until the complete screening of the initial field $E$ near the upper capacitor plate.
As shown above,  $E$ and $n_s$ are related by Eq.~(\ref{3.12ba5}).
The charges also increase the value of total electric field on the dielectric surface up to ${E_0}$ and ${{{E_0}} \mathord{\left/
		{\vphantom {{{E_0}} \varepsilon }} \right.
		\kern-\nulldelimiterspace} \varepsilon }$ below the surface (see Eq.~(\ref{3.9a5})).
The potential difference between the capacitor plates is now equal to $\Delta U = {\bar \varphi _1}\left( { + \infty } \right) - {\bar \varphi _2}\left( { - d} \right)$.
In papers (see, e.g., Refs.~\cite{ufn2011Shikin,prl1979GrimesA,jetp1975Chernikova}) it is usually considered that the electric field above liquid dielectric is zero.
If the electric field above the dielectric is absent, the potentials of the upper capacitor plate and the dielectric surface coincide.
Consequently, the voltage drop $\Delta U$ occurs only in dielectric, but not above it.
That is why the electric field inside liquid dielectric is equal to ${{\Delta U} \mathord{\left/
		{\vphantom {{\Delta U} d}} \right.
		\kern-\nulldelimiterspace} d}$.
From the other hand, as shown above, this field is equal to ${{{E_0}} \mathord{\left/
		{\vphantom {{{E_0}} \varepsilon }} \right.
		\kern-\nulldelimiterspace} \varepsilon }$ and according to Eq.~(\ref{3.9a5}) we have
\begin{eqnarray}
{{\Delta U} \mathord{\left/
		{\vphantom {{\Delta U} d}} \right.
		\kern-\nulldelimiterspace} d} = {{4\pi e{n_s}} \mathord{\left/
		{\vphantom {{4\pi e{n_s}} \varepsilon }} \right.
		\kern-\nulldelimiterspace} \varepsilon }.
\label{3.12ca5}
\end{eqnarray}
If the dielectric is liquid helium, which permittivity value is close to unit, in literature (see Refs.~\cite{prl1979GrimesA,jetp1975Chernikova}) you can often face the value of clamping field inside the capacitor equal to $E = 4\pi e{n_s}$.
In such cases we compare the values of $n_s$, obtained from our theory to the experimental value of ${n_s} = {E \mathord{\left/
		{\vphantom {E {\left( {4\pi e} \right)}}} \right.
		\kern-\nulldelimiterspace} {\left( {4\pi e} \right)}}$.
Speaking about the absence of total electric field above liquid dielectric, we can characterize this statement as approximate, but highly accurate at that.
Indeed, $\Delta U = {{4\pi e{n_s}d} \mathord{\left/
		{\vphantom {{4\pi e{n_s}d} \varepsilon }} \right.
		\kern-\nulldelimiterspace} \varepsilon } + {{T\left( {\chi \left( { + \infty } \right) - {\chi _0}} \right)} \mathord{\left/
		{\vphantom {{T\left( {\chi \left( { + \infty } \right) - {\chi _0}} \right)} e}} \right.
		\kern-\nulldelimiterspace} e}$, where the second term is the voltage drop between the upper capacitor plate and the surface of liquid dielectric.
Putting $T = 2,5K$, ${n_s} = 5 \cdot {10^8}c{m^{ - 2}}$ and $d = 0,1cm$, we obtain ${{4\pi e{n_s}d} \mathord{\left/
		{\vphantom {{4\pi e{n_s}d} \varepsilon }} \right.
		\kern-\nulldelimiterspace} \varepsilon } \approx {10^{ - 3}}V$.
And the voltage drop ${{T\left( {\chi \left( z \right) - {\chi _0}} \right)} \mathord{\left/
		{\vphantom {{T\left( {\chi \left( z \right) - {\chi _0}} \right)} e}} \right.
		\kern-\nulldelimiterspace} e}$ at the macroscopic distance from the dielectric surface, e.g., $z = 10cm$ is approximately equal to $7,9 \cdot {10^{ - 8}}V$.
This fact provides using Eq.~(\ref{3.12ca5}) with rather good accuracy.

According to Eqs.~(\ref{2.16a5}) -~(\ref{2.18a5}),~(\ref{3.9a5}),~(\ref{3.12ba5}) -~(\ref{3.14a5}), the potentials of total and external electric fields in liquid film and solid dielectric substrate are determined by the expressions:
\begin{eqnarray}
\nonumber
{\bar \varphi _2}(z) =  - \frac{E_0}{\varepsilon }\left( {z - \bar \xi } \right) + {\varphi _0},
\quad
{\bar \varphi _3}(z) =  - \frac{E_0}{{{\varepsilon _d}}}\left( {z + d} \right) + \frac{E_0}{\varepsilon }\left( {d + \bar \xi } \right) + {\varphi _0},
\quad
\bar \varphi _1^{\left( e \right)}(z) =  - E\left( {z - \bar \xi } \right) + \varphi _0^{\left( e \right)},
\\
\bar \varphi _2^{\left( e \right)}(z) =  - \frac{E}{\varepsilon }\left( {z - \bar \xi } \right) + \varphi _0^{\left( e \right)},
\quad
\bar \varphi _3^{\left( e \right)}(z) =  - \frac{E}{{{\varepsilon _d}}}\left( {z + d} \right) + \frac{E}{\varepsilon }\left( {d + \bar \xi } \right) + \varphi _0^{\left( e \right)},
\label{3.12a5}
\end{eqnarray}
where ${\bar \varphi _{1\xi }} = {\bar \varphi _{2\xi }} \equiv {\varphi _0}$, $\bar \varphi _{1\xi }^{\left( e \right)} = \bar \varphi _{2\xi }^{\left( e \right)} \equiv \varphi _0^{\left( e \right)}$.

\begin{figure}
	\begin{minipage}{.49\textwidth}
		\includegraphics[scale=0.85]{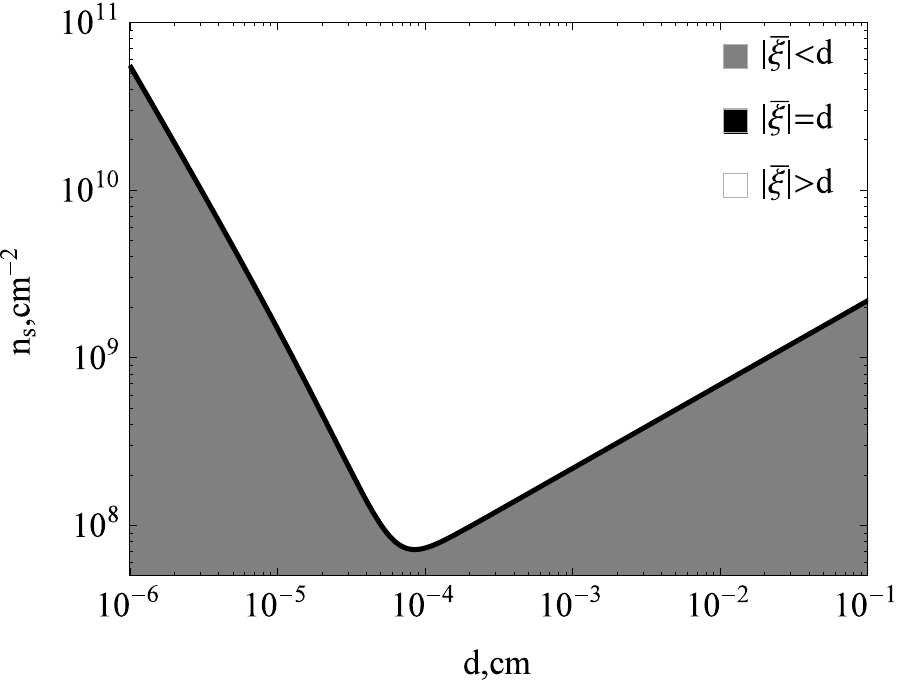}
		\caption{\label{fig:pic4aa5} Diagram of the system stability against surface deformation of liquid dielectric film in $\left\{ {d,{n_s}} \right\}$ plane.}
	\end{minipage}
	\hfill
	\begin{minipage}{.49\textwidth}
		\includegraphics[scale=0.85]{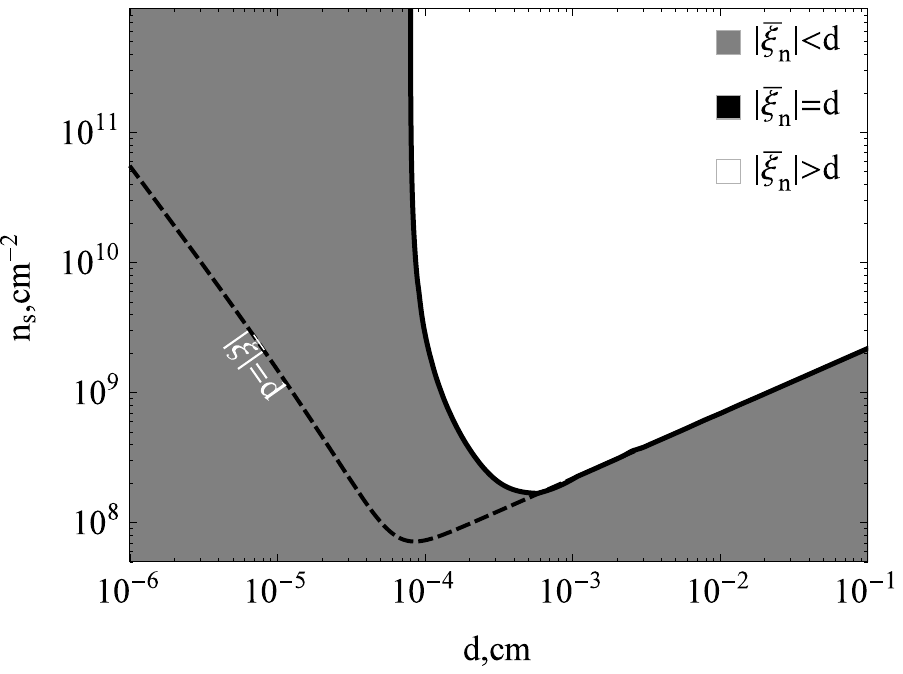}
		\caption{\label{fig:pic4ba5}   The stability $\left\{ {d,{n_s}} \right\}$ diagram with the account of ``effective'' film thickness effect Eq.~(\ref{3.13da5}).}
	\end{minipage}
\end{figure}

Basing on Eqs.~(\ref{2.15a5}),~(\ref{3.5a5}),~(\ref{3.9a5}),~(\ref{3.12ba5}),~(\ref{3.12a5}),  we obtain the value of the dielectric surface subsidence:
\begin{eqnarray}
\bar \xi  =  - \frac{{{{\left( {4\pi e{n_s}} \right)}^2}}}{{8\pi \alpha {{\left( {\kappa \left( d \right)} \right)}^2}}}\left( {1 + \frac{3}{{4\varepsilon }}} \right).
\label{3.13a5}
\end{eqnarray}
According to Eq.~(\ref{3.13a5}), in the case of charges absence the value of dielectric surface subsidence $\bar \xi $ is zero.
This value is in  good agreement with the experimental data of Ref.~\cite{pla1971CrandallW36}.
Eq.~(\ref{3.13a5}) allows imposing a natural constraint on the clamping electric field and, as a consequence, on the permissible surface electron density.
Indeed, in the system being in the equilibrium state, described by the self-consistent Eqs.~(\ref{1.1a5})~-~(\ref{1.4a5}), the absolute value of the surface subsidence of liquid dielectric film must be substantially smaller, comparing to the thickness of this film (or, at least, several times smaller):
\begin{eqnarray}
\left| {\bar \xi } \right| <  < d.
\label{3.13aa5}
\end{eqnarray}
This condition allows defining the maximum value of charges number per dielectric surface unit $n_s^m$, significantly exceeding the values under consideration:
\begin{eqnarray}
{n_s} <  < n_s^m,
\quad
n_s^m < \frac{{\kappa \left( d \right)\sqrt {{{\alpha d} \mathord{\left/
					{\vphantom {{\alpha d} {\left( {2\pi } \right)}}} \right.
					\kern-\nulldelimiterspace} {\left( {2\pi } \right)}}} }}{{e\sqrt {1 + {3 \mathord{\left/
					{\vphantom {3 {\left( {4\varepsilon } \right)}}} \right.
					\kern-\nulldelimiterspace} {\left( {4\varepsilon } \right)}}} }}.
\label{3.13ba5}
\end{eqnarray}
In the case of macroscopic values of the film thickness (the so-called massive helium case), the value of $\kappa$ is almost independent on $d$, that allows estimating the value of  $n_s^{m}$.
E.g., for the liquid helium film, having thickness $d = 0,1cm$, the value of $n_s^m \approx 2,18 \cdot {10^9}c{m^{ - 2}}$, which is comparable to the value of $n_s^{cr} \approx 2,2 \cdot {10^9}c{m^{ - 2}}$ (see Ref.~\cite{jetpl1971GorkovC}), determining the instability condition of a homogeneous electron system above the flat surface of liquid helium.
For such system the instability means the appearing of a static deformation of liquid helium surface, having a periodic structure, as the  result of forming of standing gravitational waves (see Ref.~\cite{prb1978MimaI}).
However, the value of $\kappa$  depends on film thickness $d$.
E.g., for liquid helium, this dependence has the form(see Ref.~\cite{book1989ShikinM}):
\begin{eqnarray}
\kappa \left( d \right) = \sqrt {\frac{\rho }{\alpha }\left( {g + \frac{{{g_0}{d_v}}}{{{d^4}\left( {d + {d_v}} \right)}}\left( {3 + \frac{d}{{d + {d_v}}}} \right)} \right)},
\label{3.13ca5}
\end{eqnarray}
where ${d_v} = 1,65 \cdot {10^{ - 5}}cm$ and ${g_0} = 2,2 \cdot {10^{ - 14}}c{m^5} \cdot {s^{ - 2}}$.
So, according to Eq.~(\ref{3.13a5}), in some cases this dependency should be taken into account during the analysis of the stability condition Eq.~(\ref{3.13aa5}).
The diagram, illustrating the satisfaction of the stability condition Eq.~(\ref{3.13aa5}) of the system, is presented on Fig.~\ref{fig:pic4aa5} in $\left\{ {d,{n_s}} \right\}$ plane.
Fig.~\ref{fig:pic4aa5} takes into account Eq.~(\ref{3.13ca5}).
On this figure gray color marks the region, where ${{\left| \xi  \right|} \mathord{\left/
		{\vphantom {{\left| \xi  \right|} d}} \right.
		\kern-\nulldelimiterspace} d} < 1$.
In other words, according to our theory, this region is the stability region of the system, relatively to the surface deformations, as the result of the gas charges pressure.
The black line, separating the gray and white regions, corresponds to the equality $\left| {\bar \xi } \right| = d$.
Fig.~\ref{fig:pic4aa5} shows that helium is considered to be ``massive'', if $d > 5 \cdot {10^{ - 2}}cm$.
Let us also note, that Fig.~\ref{fig:pic4aa5} has a good agreement with the data of Ref.~\cite{prb1984Peeters}.
In this paper the author obtained the relation between the thickness of dielectric film and  maximum available value of $n_s$, at which the surface of this film stays flat.
This relation was obtained by solving the problem on the stability of small oscillations in such system.

The decreasing of film thickness decreases the contribution of gravitational force in $\kappa$ and increases the contribution of Van der Waals forces.
This competitive process between gravitational and Van der Waals forces ends at $d \sim {d_v}$, when the gravitational forces, acting on the atoms of liquid dielectric, becomes negligibly small, comparing to Van der Waals forces.

However, the decreasing of film thickness often results in the situation, where the orders of $\left| {\bar \xi } \right|$ and $d$ are comparable.
In this case in Eqs.~(\ref{2.13a5}) and~(\ref{3.13a5}), we have to substitute the film thickness $d$ by the ``effective'' thickness $d - \left|  {\bar \xi }  \right|$.
This substitution results in the following equation for $\bar \xi$ calculation
\begin{eqnarray}
\bar \xi  =  - \frac{{{{\left( {4\pi e{n_s}} \right)}^2}}}{{8\pi \alpha {{\left( {\kappa \left( {d - \left| {\bar \xi } \right|} \right)} \right)}^2}}}\left( {1 + \frac{3}{{4\varepsilon }}} \right).
\label{3.13da5}
\end{eqnarray}
Taking into account Eq.~(\ref{3.13ca5}), it is easily seen, that Eq.~(\ref{3.13da5}) has no analytical solution for $\bar \xi$.
The numeric solution of Eq.~(\ref{3.13da5}) ${\bar \xi _n}$ provides new region $\left| {{{\bar \xi }_n}} \right| < d$ of the system stability against the deformations of liquid helium surface.
The stability region $\left| {{{\bar \xi }_n}} \right| < d$ is marked on Fig.~\ref{fig:pic4ba5} by gray color.
And the region $d > \left| {\bar \xi } \right|$ is located below the dashed line.
Fig.~\ref{fig:pic4ba5} shows, that taking into account the effective film thickness is significant for ``thin'' helium films in the range of film thickness $d < 5 \cdot {10^{ - 4}}cm$.
In massive helium case this effect vanishes.
In the case of thin helium films the region of available values on $\left\{ {{n_s},d} \right\}$ plane is much wider than in the case of ignoring this effect (see Fig.~\ref{fig:pic4aa5}).

\begin{figure}
	\begin{minipage}{.49\textwidth}
		\includegraphics[scale=0.85]{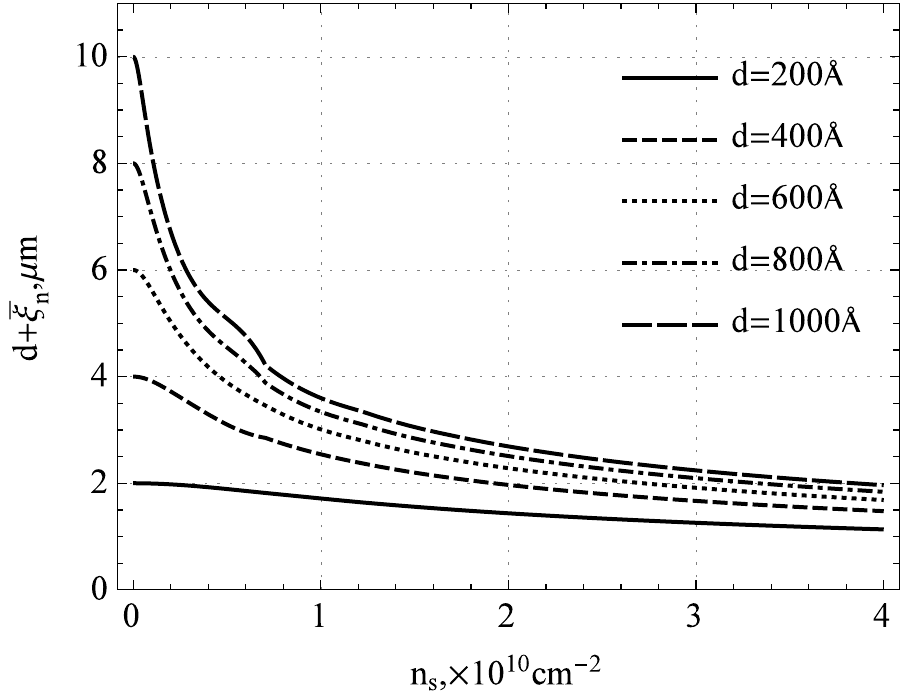}
		\caption{\label{fig:pic5a5} The dependence of effective film thickness of liquid dielectric on $n_s$.}
	\end{minipage}
	\hfill
	\begin{minipage}{.49\textwidth}
		\includegraphics[scale=0.85]{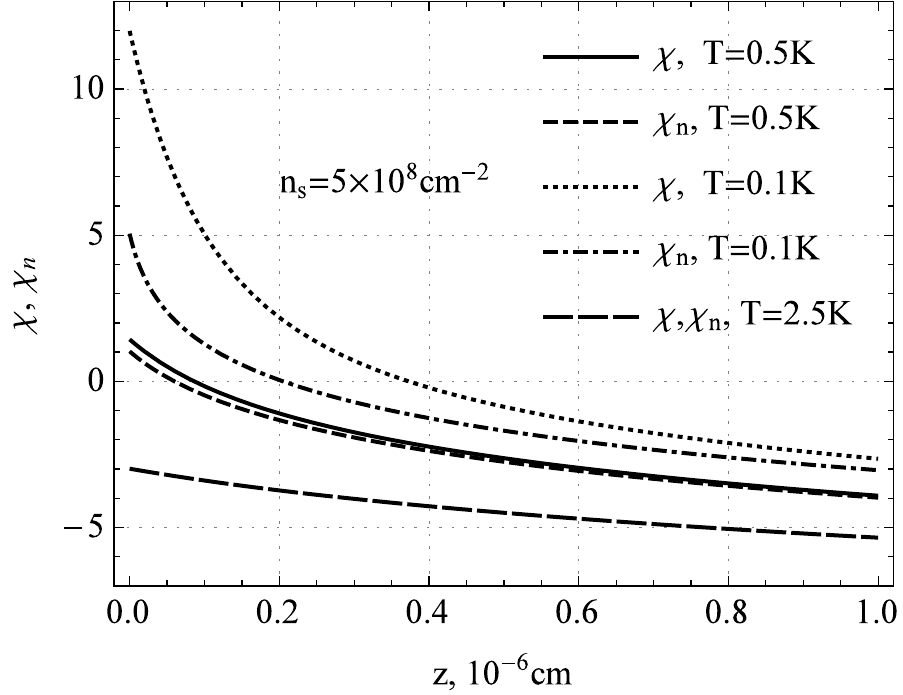}
		\caption{\label{fig:pic6aa5}  Electrochemical potential $\chi \left( z \right)$ in general and non-degenerate cases for three pairs of $T$ and $n_s$ values.}
	\end{minipage}
\end{figure}

Let us also note one more important fact.
In the case of thin enough films, having thickness $d \sim {10^{ - 6}} \div {10^{ - 5}}cm$ (if the clamping field is absent), and high enough values of ${n_s} > 5 \cdot {10^{10}}c{m^{ - 2}}$, the effective film thickness $d - \left| {{{\bar \xi }_n}} \right|$ is almost independent on $d$.
This fact is illustrated on Fig.~\ref{fig:pic5a5}, which is in good agreement with the data of Refs.~\cite{prb1990HuD,prl1984EtzGIL}.
If ${n_s} > 4 \cdot {10^{11}}c{m^{ - 2}}$, the effective dielectric film thickness reaches a value of about 50 angstroms.
According to Ref.~\cite{prb1990HuD}, this fact can lead to the electron tunneling through dielectric film towards the metal substrate.
However, the study of this effect goes out of the scope of the present paper.
In fact, it aims developing the quasi-classical description of a system of charges in terms of Wigner distribution function Eq.~(\ref{1.2a5}), simultaneously depending on ${\bf{r}}$ and ${\bf{p}}$.
So, in this approach, the description of quantum mechanical tunneling effect is not possible.
However, using the Eq.~(\ref{3.13da5}) provides obtaining numerical estimates, comparable with the results of Ref.~\cite{prl1984EtzGIL}.
E.g., for the thickness of helium film $d < {10^{ - 6}}cm$ on a metal substrate, we obtain $n_s^m \approx 0,7 \cdot {10^{11}}c{m^{ - 2}}$.

\begin{figure}
	\begin{minipage}{.49\textwidth}
		\includegraphics[scale=0.85]{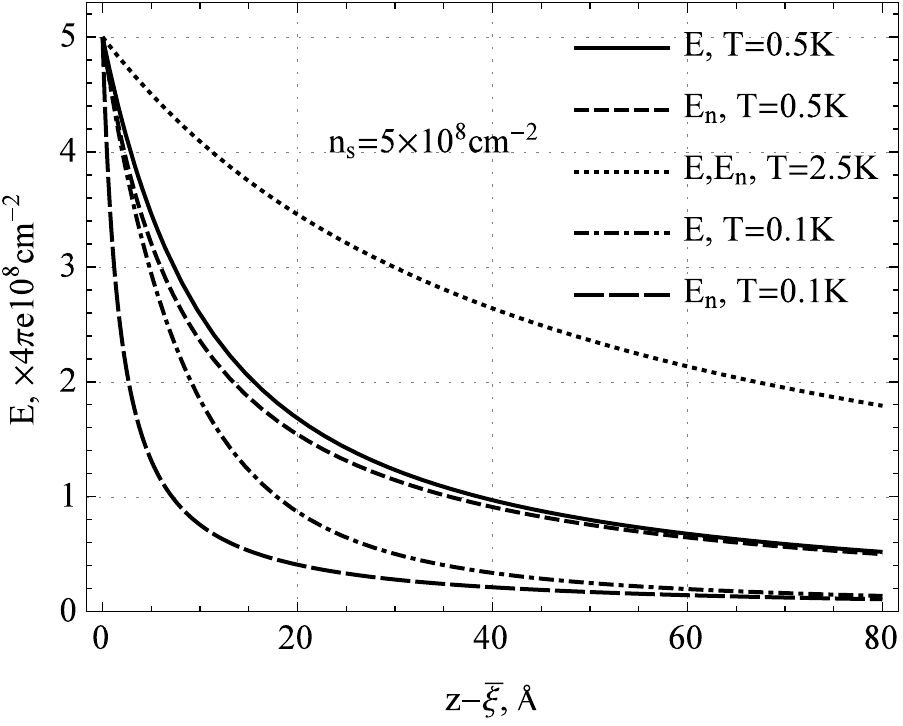}
		\caption{\label{fig:pic6ba5} Electric field in ``1''~region ${E_1}\left( z \right)$ in general and non-degenerate cases for three pairs of $T$ and $n_s$ values.}
	\end{minipage}
	\hfill
	\begin{minipage}{.49\textwidth}
		\includegraphics[scale=0.88]{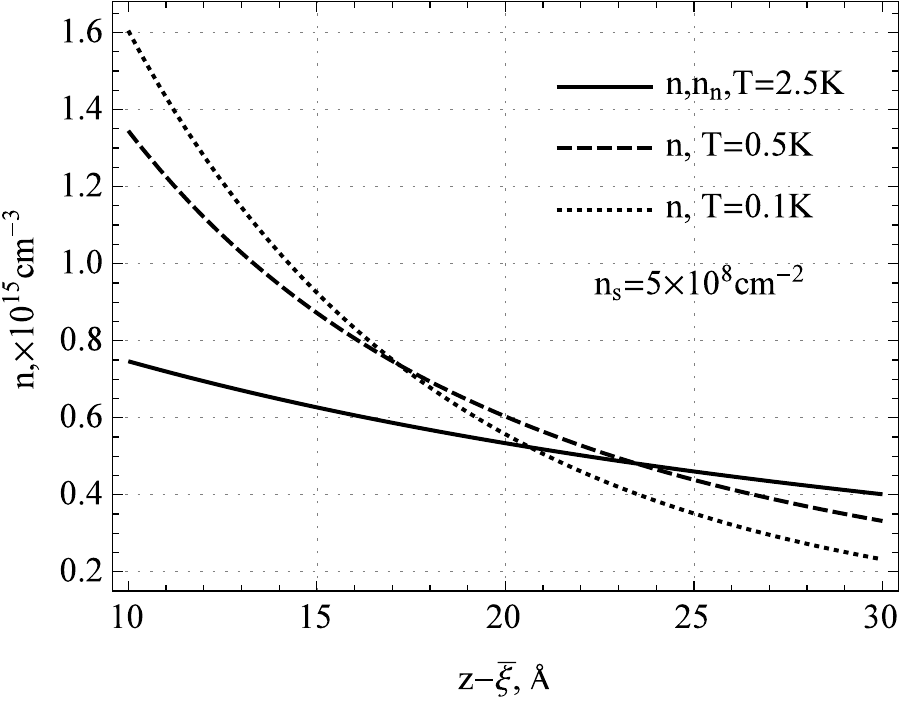}
		\caption{\label{fig:pic7aa5}  Particle volume density $n\left( z \right)$ in general and non-degenerate cases for three values of $T$.}
	\end{minipage}
\end{figure}

The dependence $\chi \left( z \right)$, related with ${\bar \varphi _1}\left( z \right)$ by Eqs.~(\ref{3.2a5}),~(\ref{3.4a5}) is obtained by the numeric integration of Eq.~(\ref{3.4a5}), or
\begin{eqnarray}
\frac{{{a_0}}}{{{2^{{5 \mathord{\left/
						{\vphantom {5 4}} \right.
						\kern-\nulldelimiterspace} 4}}}}}{\left( {\frac{{\pi {e^2}}}{{T{a_0}}}} \right)^{{1 \mathord{\left/
				{\vphantom {1 4}} \right.
				\kern-\nulldelimiterspace} 4}}}\int\limits_{{\chi _0}}^\chi  {\frac{{d\chi '}}{{\sqrt { - L{i_{{5 \mathord{\left/
								{\vphantom {5 2}} \right.
								\kern-\nulldelimiterspace} 2}}}\left( { - {e^{\chi '}}} \right)} }}}  = \bar \xi  - z.
\label{3.14a5}
\end{eqnarray}
Fig.~\ref{fig:pic6aa5} shows the comparison between $\chi \left( z \right)$ and ${\chi _n}\left( z \right)$ for exact values of $T$ and $n_s$.
${\chi _n}\left( z \right)$ is the non-degenerate analog of $\chi \left( z \right)$ function, obtained in Ref.~\cite{jps2015SlyusarenkoL}:
\begin{eqnarray}
{\chi _n}\left( z \right) = {\chi _{n0}} - 2\ln \left( {{{1 + \left( {z - \bar \xi } \right)} \mathord{\left/
			{\vphantom {{1 + \left( {z - \bar \xi } \right)} {\left( {2{z_0}} \right)}}} \right.
			\kern-\nulldelimiterspace} {\left( {2{z_0}} \right)}}} \right),
\quad
{\chi _{n0}} = \ln \left( {\frac{{{n_s}a_0^3}}{{\sqrt 2 {z_0}}}{{\left( {\frac{{\pi {e^2}}}{{T{a_0}}}} \right)}^{{3 \mathord{\left/
					{\vphantom {3 2}} \right.
					\kern-\nulldelimiterspace} 2}}}} \right),
\quad
{z_0} = {T \mathord{\left/
		{\vphantom {T {\left( {e{E_0}} \right)}}} \right.
		\kern-\nulldelimiterspace} {\left( {e{E_0}} \right)}}.
\label{3.15a5}
\end{eqnarray}
Three selection cases of the specific pairs of values of $T$ and $n_s$ are chosen, basing on the principle of satisfying the non-degeneracy condition of electron gas (see Ref.~\cite{jps2015SlyusarenkoL}):
\begin{eqnarray}
{2^{3/2}}n_s^2a_0^4{\left( {\frac{{\pi {e^2}}}{{T{a_0}}}} \right)^{{5 \mathord{\left/
				{\vphantom {5 2}} \right.
				\kern-\nulldelimiterspace} 2}}} <  < 1.
\label{3.16a5}
\end{eqnarray}
In first case $T = 2,5K$ and ${n_s} = 5 \cdot {10^8}c{m^{ - 2}}$, and the gas of charges is non-degenerate, as ${2^{3/2}}n_s^2a_0^4{\left( {{{\pi {e^2}} \mathord{\left/
				{\vphantom {{\pi {e^2}} {\left( {T{a_0}} \right)}}} \right.
				\kern-\nulldelimiterspace} {\left( {T{a_0}} \right)}}} \right)^{{5 \mathord{\left/
				{\vphantom {5 2}} \right. \kern-\nulldelimiterspace} 2}}} \approx 0.05$.
That is why, in this case we observe the practical coincidence of $\chi \left( z \right)$ and ${\chi _n}\left( z \right)$ curves.
The second case is $T = 0,5K$ and ${n_s} = 5 \cdot {10^8}c{m^{ - 2}}$.
In this case the non-degeneracy condition Eq.~(\ref{3.16a5}) breaks (${2^{3/2}}n_s^2a_0^4{\left( {{{\pi {e^2}} \mathord{\left/
				{\vphantom {{\pi {e^2}} {\left( {T{a_0}} \right)}}} \right.
				\kern-\nulldelimiterspace} {\left( {T{a_0}} \right)}}} \right)^{{5 \mathord{\left/
				{\vphantom {5 2}} \right.
				\kern-\nulldelimiterspace} 2}}} \approx 2,79$) and it is observable on Fig.~\ref{fig:pic6aa5}.
If the distance from the dielectric surface ${z - \bar \xi }$ increases, the gas density Eq.~(\ref{3.6a5}) decreases (see Fig.~\ref{fig:pic7aa5}), so as the distance between $\chi \left( z \right)$ and ${\chi _n}\left( z \right)$ curves does.
If $z - \bar \xi  = 10{z_0} \approx 4,45 \cdot {10^{ - 7}}cm$, the ratio ${\chi  \mathord{\left/
		{\vphantom {\chi  {{\chi _n}}}} \right.
		\kern-\nulldelimiterspace} {{\chi _n}}} \approx 0,95$.
With further increasing of the distance from dielectric surface the gas can be considered as non-degenerate with 95 percent accuracy.
This explains the practical coincidence of $\chi \left( z \right)$ and ${\chi _n}\left( z \right)$ curves in this range of ${z - \bar \xi }$.
In the third case, $T = 0,1K$ and ${n_s} = 5 \cdot {10^8}c{m^{ - 2}}$, the condition Eq.~(\ref{3.16a5}) breaks dramatically (${2^{3/2}}n_s^2a_0^4{\left( {{{\pi {e^2}} \mathord{\left/
				{\vphantom {{\pi {e^2}} {\left( {T{a_0}} \right)}}} \right.
				\kern-\nulldelimiterspace} {\left( {T{a_0}} \right)}}} \right)^{{5 \mathord{\left/
				{\vphantom {5 2}} \right.
				\kern-\nulldelimiterspace} 2}}} \approx 156,18$), and the gas of charges cannot be considered as non-degenerate.
However, unlikely the previous case, the gas of charges can be considered as non-degenerate at distances greater than $z - \bar \xi  = 200{z_0} \approx 1,78 \cdot {10^{ - 6}}cm$ with not less than 95 percent accuracy.

Let us note, that the value of electric field ${E_1}\left( z \right)$ is obtained by the differentiation ${E_1} =  - \chi '\left( z \right){T \mathord{\left/
		{\vphantom {T e}} \right.
		\kern-\nulldelimiterspace} e}$.
This dependence is presented on Fig.~\ref{fig:pic6ba5} in three mentioned above cases together with the dependence ${E_n}\left( z \right)$ of non-degenerate gas, obtained in Ref.~\cite{jps2015SlyusarenkoL}:
\begin{eqnarray}
{E_n}\left( z \right) = {{{E_0}} \mathord{\left/
		{\vphantom {{{E_0}} {\left( {{{1 + \left( {z - \bar \xi } \right)} \mathord{\left/
								{\vphantom {{1 + \left( {z - \bar \xi } \right)} {\left( {2{z_0}} \right)}}} \right.
								\kern-\nulldelimiterspace} {\left( {2{z_0}} \right)}}} \right)}}} \right.
		\kern-\nulldelimiterspace} {\left( {{{1 + \left( {z - \bar \xi } \right)} \mathord{\left/
					{\vphantom {{1 + \left( {z - \bar \xi } \right)} {\left( {2{z_0}} \right)}}} \right.
					\kern-\nulldelimiterspace} {\left( {2{z_0}} \right)}}} \right)}}.
\label{3.18a5}
\end{eqnarray}
Fig.~\ref{fig:pic6ba5} shows, that the decreasing rate of electric field with the distance ${z - \bar \xi }$ increasing is defined by the extent of gas non-degeneracy.
The closer gas is to the non-degeneracy state, the slower the decreasing of electric field value with ${z - \bar \xi }$ increasing is.
On Fig.~\ref{fig:pic6ba5} all curves are starting from one point at $z = \bar \xi$, because all of them have the same value of ${n_s} = 5 \cdot {10^8}c{m^{ - 2}}$, corresponding to the electric field value Eq.~(\ref{3.9a5}).

Basing on Eq.~(\ref{3.14a5}), $\chi \left( z \right)$ dependency is obtained, which allows obtaining $n\left( z \right)$, using Eq.~(\ref{3.6a5}).
Fig.~\ref{fig:pic7aa5} shows $n\left( z \right)$ curve in three mentioned above cases of pair values $T$ and $n_s$.
In first case $T = 2,5K$ and ${n_s} = 5 \cdot {10^8}c{m^{ - 2}}$ the gas of charges is close to the non-degenerate state.
So, in this case the $n\left( z \right)$  curve practically coincides the ${n_n}\left( z \right)$ curve of non-degenerate gas, obtained in Refs.~\cite{jps2015SlyusarenkoL,jetp1975Chernikova}:
\begin{eqnarray}
{n_n}\left( z \right) = \frac{{{n_s}}}{{2{z_0}}}{\left( {1 + {{\left( {z - \bar \xi } \right)} \mathord{\left/
				{\vphantom {{\left( {z - \bar \xi } \right)} {\left( {2{z_0}} \right)}}} \right.
				\kern-\nulldelimiterspace} {\left( {2{z_0}} \right)}}} \right)^{ - 2}}.
\label{3.17a5}
\end{eqnarray}
Let us note, that the closer gas is to the non-degeneracy state, the slower its density decreases with distance from the dielectric surface growth.
A particular interest is a gas, being in state, that is close to degeneracy.
This state is realized at sufficiently high density and low temperature range.
In this case an important role in the inter-particle interaction can play the exchange processes.
However, the detailed research of this effect is out of the present paper scope.

Using the obtained above density function $n\left( z \right)$, we can estimate certain typical distances of the considered system.
Let us introduce the distance from the dielectric surface, characterizing the volume, containing the major part of system charges.
With this purpose we introduce the following function
\begin{eqnarray}
\Delta \left( z \right) = \frac{1}{{{n_s}}}\int\limits_{\bar \xi }^z {n\left( x \right)} dx,
\label{3.19a5}
\end{eqnarray}
giving the relative percentage of charges, located between the dielectric surface and the distance $z$ above it.
\begin{figure}% width=0.25\textwidth,height=0.25\textwidth
	\centering
	\includegraphics[scale=1.1]{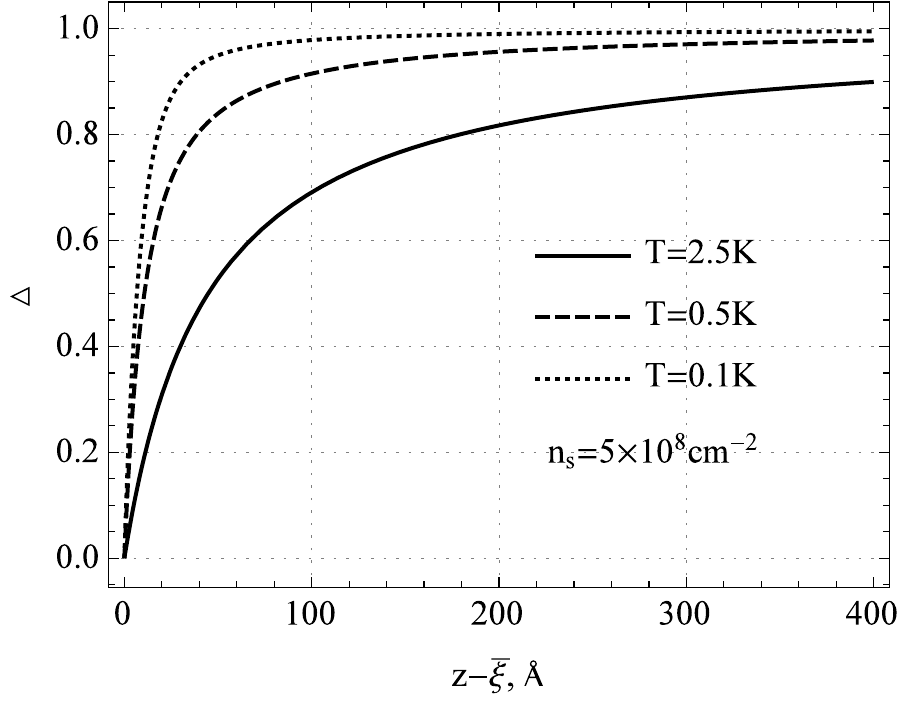}
	\caption{\label{fig:pic7ba5}  %Relative percentage of charges $\Delta \left( z \right)$, located in $\left( {\bar \xi ,\;z} \right)$ region, for three values of $T$.
		$\Delta \left( z \right)$  for different temperature values.}
\end{figure}
Fig.~\ref{fig:pic7ba5} shows $\Delta \left( z \right)$ curves for different temperature values and fixed density value ${n_s} = 5 \cdot {10^8}c{m^{ - 2}}$.
Let us introduce ${z_m}$ distance by the following definition $\Delta \left( {{z_m}} \right) \approx 0,95$.
In other words, the number of charges, located in the range ${z_m} \ge z \ge \bar \xi$, is equal to 95 percent of the total number of charges in the system.
Taking this into account, we can consider ${z_m}$ as approximate or ``effective'' boundary for the gas of charges.
So, for the temperature $T = 2,5K$, we have ${z_m} = 40{z_0} \approx 8,9 \cdot {10^{ - 6}}cm$.
In the case of $T = 0,5K$, ${z_m} = 40{z_0} \approx 1,78 \cdot {10^{ - 6}}cm$.
And in the case of $T = 0,1K$, ${z_m} = 57{z_0} \approx 5,1 \cdot {10^{ - 7}}cm$.
As expected, we can see, that in the case of fixed total number of charges in the gas, its effective boundary decreases, if the temperature decreases.
Let us emphasize, that the distance $z_m$ has the same order of value, as the localization distance above helium surface, that a single electron in the ground state has (see Ref.~\cite{jetp1970Shikin,ssc1978HippolitoFF}).
The problem on obtaining such localization distance can be reduced to the problem on obtaining the average electron distance from the nucleus in hydrogen atom, being in the ground state.
The authors of Ref.~\cite{jetpl1971GorkovC} pointed out, that in the range of ${n_s} \sim {10^8} \div {10^9}c{m^{ - 2}}$ the mean distance between charges was one or two orders greater than the electron localization above helium surface.
For this reason, they considered the gas of electrons as two-dimensional.

In present paper we consider the case, where the value of introduced above distance $z_m$ is also more than one order less than $n_s^{{{ - 1} \mathord{\left/
			{\vphantom {{ - 1} 2}} \right.
			\kern-\nulldelimiterspace} 2}}$.
At first glance this fact justifies considering the studied system as ``quasi-two-dimensional''. Indeed, at certain values of $T$ and $n_s$, the mean distance between the charges projections on the flat dielectric surface, which is proportional to $n_s^{ - {1 \mathord{\left/
			{\vphantom {1 2}} \right.
			\kern-\nulldelimiterspace} 2}}$, can be two orders greater than $z_m$.
For this reason, charges can be considered being located almost in one plane with the accuracy up to the small value ${z_m}n_s^{{1 \mathord{\left/
			{\vphantom {1 2}} \right.
			\kern-\nulldelimiterspace} 2}} \ll 1$.
However, in most general case the inequality ${z_m}n_s^{{1 \mathord{\left/
			{\vphantom {1 2}} \right.
			\kern-\nulldelimiterspace} 2}} \ll 1$ breaks.
And even if this inequality takes place, it can not serve as a justification for the system two-dimensional consideration.
Let us prove the last statement by calculating  the mean distance from flat dielectric surface to charges, located in volume between two planes $z = \bar \xi$ and $z = {z_m}$.
The probability of charge location in the range $\left( {z,\;z + dz} \right)$ from the dielectric surface is equal to ${{n\left( z \right)dz} \mathord{\left/
		{\vphantom {{n\left( z \right)dz} {{n_s}}}} \right.
		\kern-\nulldelimiterspace} {{n_s}}}$.
Then, according to Eqs.~(\ref{3.4a5}),~(\ref{3.6a5}), and~(\ref{3.8a5}) the mean distance from charge to the dielectric surface has the following form:
\begin{eqnarray}
\left\langle {z - \bar \xi } \right\rangle  = n_s^{ - 1}\int\limits_{\bar \xi }^{{z_m}} {dz} n\left( z \right)\left( {z - \bar \xi } \right) = {z_0}\left. \chi  \right|_{{z_m}}^{\bar \xi }.
\label{3.20a5}
\end{eqnarray}
Let us estimate this value in three mentioned above cases of $T$ and $n_s$ values.
In each pair ${n_s} = 5 \cdot {10^8}c{m^{ - 2}}$, and for $T = 2,5K$ we have $\left\langle {z - \bar \xi } \right\rangle  \approx 4{z_0} = 0,11 \cdot {z_m} \approx 0,9 \cdot {10^{ - 6}}cm$, for $T = 0,5K$ we obtain  $\left\langle {z - \bar \xi } \right\rangle  \approx 4,5{z_0} = 0,15{z_m} \approx 2 \cdot {10^{ - 7}}cm$, and for $T = 0,1K$ - $\left\langle {z - \bar \xi } \right\rangle  \approx 10{z_0} = 0,18{z_m} \approx 0,9 \cdot {10^{ - 7}}cm$.
So, in the considered region of $\left\{ {T,{n_s}} \right\}$ plane, in the volume between $z = \bar \xi$ and $z = {z_m}$ planes, the mean distance from charges to the dielectric surface $z = \bar \xi$ is 5-10 times less than than the typical distance $z_m$ of charges localization above the surface of dielectric film.
This is the main obstacle for considering the studied quasi-neutral system of charges above liquid dielectric, as 2D system.
Taking into account the inequality ${z_m}n_s^{{1 \mathord{\left/
			{\vphantom {1 2}} \right.
			\kern-\nulldelimiterspace} 2}} \ll 1$ and the above estimates, the mean distance between charges along $z$ axis $\left\langle z \right\rangle$ is small, comparing to the mean distance between them in $\left\{ {x,y} \right\}$ plane, which is proportional to $\sqrt{n_s^{ - 1}}$.
So, if $\left\langle z \right\rangle  \ll \sqrt {n_s^{ - 1}}$, the mean distance between charges $l$  also has the order of $\sqrt {n_s^{ - 1}}$, according to the estimation $l \sim \sqrt {n_s^{ - 1} + {{\left\langle z \right\rangle }^2}}  \approx \sqrt {n_s^{ - 1}}$.
This fact allows us obtaining the applicability condition for the quasi-classical approach, used in the present paper.

To make the corresponding estimations, we calculate the mean thermal de Broglie wavelength $\left\langle \lambda  \right\rangle  \sim {\hbar  \mathord{\left/
		{\vphantom {\hbar  {\sqrt {\left\langle {{p^2}} \right\rangle } }}} \right.
		\kern-\nulldelimiterspace} {\sqrt {\left\langle {{p^2}} \right\rangle } }}$ of charges above dielectric surface.
Taking into account Eqs.~(\ref{1.2a5}),~(\ref{3.4a5})~-~(\ref{3.6a5}), the mean value of squared momentum has the following form:
\begin{eqnarray}
\nonumber
\left\langle {{p^2}} \right\rangle  = {{\int {{d^3}r{d^3}p} {f_{\bf{p}}}\left( {\bf{r}} \right){p^2}} \mathord{\left/
		{\vphantom {{\int {{d^3}r{d^3}p} {f_{\bf{p}}}\left( {\bf{r}} \right){p^2}} {\int {{d^3}r{d^3}p} {f_{\bf{p}}}\left( {\bf{r}} \right)}}} \right.
		\kern-\nulldelimiterspace} {\int {{d^3}r{d^3}p} {f_{\bf{p}}}\left( {\bf{r}} \right)}}
= {\left( {\frac{{T{a_0}}}{{\pi {e^2}}}} \right)^{\frac{5}{4}}}\frac{{3mT}}{{{2^{{7 \mathord{\left/
						{\vphantom {7 4}} \right.
						\kern-\nulldelimiterspace} 4}}}a_0^2{n_s}}}\int\limits_{ - \infty }^{{\chi _0}} {d\chi {{\left( { - L{i_{{5 \mathord{\left/
								{\vphantom {5 2}} \right.
								\kern-\nulldelimiterspace} 2}}}\left( { - {e^\chi }} \right)} \right)}^{\frac{1}{2}}}}.
\end{eqnarray}
This equation allows estimating the thermal de Broglie wavelength $\left\langle \lambda  \right\rangle$:
\begin{eqnarray}
\nonumber
\left\langle \lambda  \right\rangle  \sim {\hbar  \mathord{\left/
		{\vphantom {\hbar  {\sqrt {\left\langle {{p^2}} \right\rangle } }}} \right.
		\kern-\nulldelimiterspace} {\sqrt {\left\langle {{p^2}} \right\rangle } }} = a_0^2\sqrt {{n_s}} {\left( {\frac{{\pi {e^2}}}{{T{a_0}}}} \right)^{{9 \mathord{\left/
				{\vphantom {9 8}} \right.
				\kern-\nulldelimiterspace} 8}}}
 \sqrt {\frac{{{2^{{7 \mathord{\left/
							{\vphantom {7 4}} \right.
							\kern-\nulldelimiterspace} 4}}}}}{{3\pi }}} {\left( {\int\limits_{ - \infty }^{{\chi _0}} {d\chi {{\left( { - L{i_{{5 \mathord{\left/
										{\vphantom {5 2}} \right.
										\kern-\nulldelimiterspace} 2}}}\left( { - {e^\chi }} \right)} \right)}^{{1 \mathord{\left/
							{\vphantom {1 2}} \right.
							\kern-\nulldelimiterspace} 2}}}} } \right)^{ - {1 \mathord{\left/
				{\vphantom {1 2}} \right.
				\kern-\nulldelimiterspace} 2}}}.
\end{eqnarray}
The numeric calculation of the last equation shows, that in the range of ${10^8}c{m^{ - 2}} < {n_s} < 2 \cdot {10^9}c{m^{ - 2}}$ and $0.1K < T < 5K$, that the order of $\left\langle \lambda  \right\rangle$ value is ${10^{ - 6}}cm$.
This is two orders less, than the mean inter-particle distance, proportional to $n_s^{ - {1 \mathord{\left/
			{\vphantom {1 2}} \right.
			\kern-\nulldelimiterspace} 2}}$ (see above).
This fact allows solving the problems of present paper in terms of quasi-classical approach, neglecting such quantum type of inter-particle interaction, as exchange type.
This type of interaction can become significant near the degenerate state of the gas of charges.
This state is achieved by decreasing temperature and increasing density of charges.
As far as experimental realization of such degenerate states is concerned, it faces difficulties in reaching sufficiently low temperature range and the high density range is limited by the stability criterion Eq.~(\ref{3.13aa5}),~(\ref{3.13ba5}) and Fig.~\ref{fig:pic4ba5}.
However, decreasing the dielectric film thickness up to the thin films region, results in reaching the permitted density region, which is several orders greater than in massive dielectric case.
The description of such situation goes out of the present paper scope.
But the preliminary calculations show the availability of the theory modification in the case of degenerate gas of charges above liquid dielectric surface.
The motivation for such description comes from the experiments with thin films~\cite{prl1984EtzGIL} and theoretical papers, based on other theoretical approaches~\cite{prb1984Peeters,prb1981IkeziP,ltp1986Tatarskii}.

Eqs.~(\ref{3.9a5})~-~(\ref{3.14a5}) are the solution for the problem on obtaining the distribution of density and electric field in the quasi-neutral system of charges above liquid dielectric film in external clamping field.
These results are used in next section, devoted to the research of the phase transition, concerned with the forming of spatially-periodic states in the system.

\section{Critical parameters of the phase transition to a spatially periodic state of the system}

The starting point in the research of critical parameters of the phase transition with forming of spatially periodic structures of dimple type is Eq.~(\ref{2.21a5}).
Let us rewrite the first equation in Eq.~(\ref{2.21a5}) in the following form:
\begin{eqnarray}
\frac{{{\partial ^2}\tilde \varphi _1^{\left( 1 \right)}}}{{\partial {z^2}}} = \left( {q_0^2 - 2\sqrt 2 \frac{{L{i_{{1 \mathord{\left/
							{\vphantom {1 2}} \right.
							\kern-\nulldelimiterspace} 2}}}\left( { - {e^\chi }} \right)}}{{a_0^2}}{{\left( {\frac{{T{a_0}}}{{\pi {e^2}}}} \right)}^{{1 \mathord{\left/
					{\vphantom {1 2}} \right.
					\kern-\nulldelimiterspace} 2}}}} \right)\tilde \varphi _1^{\left( 1 \right)},
\quad
\label{4.1a5}
\end{eqnarray}
where, according to Eqs.~(\ref{3.2a5}),~(\ref{3.4a5})~-~(\ref{3.6a5}), we take into account the following expression
\begin{eqnarray}
\frac{{\partial n}}{{\partial \mu }} =  - \frac{{L{i_{{1 \mathord{\left/
						{\vphantom {1 2}} \right.
						\kern-\nulldelimiterspace} 2}}}\left( { - {e^\chi }} \right)}}{{\sqrt 2 \pi a_0^2{e^2}}}{\left( {\frac{{T{a_0}}}{{\pi {e^2}}}} \right)^{{1 \mathord{\left/
				{\vphantom {1 2}} \right.
				\kern-\nulldelimiterspace} 2}}}.
\label{4.1aa5}
\end{eqnarray}
Using Eq.~(\ref{3.4a5}), we reduce the derivatives on $z$ to the derivatives on $\chi$ and consider the case:
\begin{eqnarray}
\frac{{2\sqrt 2 }}{{q_0^2a_0^2}}{\left( {\frac{{T{a_0}}}{{\pi {e^2}}}} \right)^{{1 \mathord{\left/
				{\vphantom {1 2}} \right.
				\kern-\nulldelimiterspace} 2}}}\left| {L{i_{{1 \mathord{\left/
					{\vphantom {1 2}} \right.
					\kern-\nulldelimiterspace} 2}}}\left( { - {e^\chi }} \right)} \right| >  > 1,
\label{4.2a5}
\end{eqnarray}
which significantly simplifies the solution of Eq.~(\ref{4.1a5}).
Let us estimate the distance from the dielectric surface, where the inequality Eq.~(\ref{4.2a5}) starts breaking.
At $T = 5K$ and ${n_s} = {10^8}c{m^{ - 2}}$ and maximum possible $q_0^{} \approx 3 \cdot {10^4}c{m^{ - 1}}$ at the given $n_s$ (corresponding to the case of one charge in the lattice node), condition Eq.~(\ref{4.2a5}) takes place in the range $z - \bar \xi  < {10^{ - 5}}cm$.
Further increasing of $z$ results in the condition Eq.~(\ref{4.2a5}) breaking.
However, increasing the value of $z$ also results in the gas density decreasing and the gas becomes closer to the non-degeneracy state~\cite{jps2015SlyusarenkoL}.
According to Eq.~(\ref{2.21a5}), the critical curve is defined by the solution of Eq.~(\ref{4.1a5}) at $z = \bar \xi$, where the condition Eq.~(\ref{4.2a5}) takes place very well.
For this reason let us solve Eq.~(\ref{4.1a5}) in the neighborhood of $z = \bar \xi$ point.

\begin{figure}
	\centering
	\includegraphics[scale=1.1]{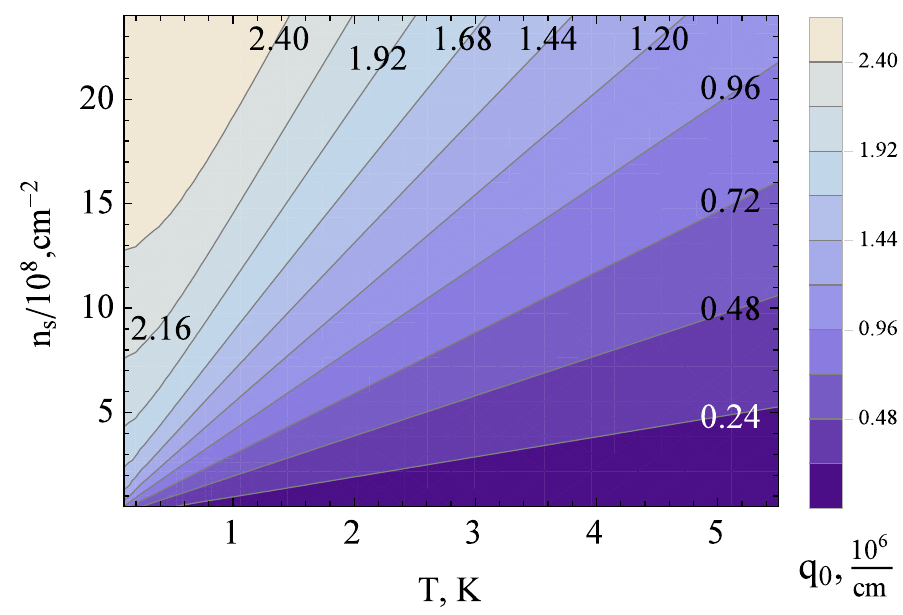}
	\caption{\label{fig:pic8a5}  Dependence of maximum available value of reciprocal lattice vector $q_0$on $T$ and $n_s$.}
\end{figure}

Fig.~\ref{fig:pic8a5} shows the dependence ${q_0}\left( {T,{n_s}} \right)$, below which the condition Eq.~(\ref{4.2a5}) takes place for the reciprocal lattice vectors ${q_0}$  and $2{q_0}$:
\begin{eqnarray}
{q_0}\left( {T,{n_s}} \right) = \frac{{\sqrt {\left| {L{i_{{1 \mathord{\left/
								{\vphantom {1 2}} \right.
								\kern-\nulldelimiterspace} 2}}}\left( { - {e^{{\chi _0}}}} \right)} \right|} }}{{{2^{{3 \mathord{\left/
						{\vphantom {3 4}} \right.
						\kern-\nulldelimiterspace} 4}}}{5^{{1 \mathord{\left/
						{\vphantom {1 2}} \right.
						\kern-\nulldelimiterspace} 2}}}a_0^{}}}{\left( {\frac{{T{a_0}}}{{\pi {e^2}}}} \right)^{{1 \mathord{\left/
				{\vphantom {1 4}} \right.
				\kern-\nulldelimiterspace} 4}}}.
\end{eqnarray}
The necessity of $2q_0$ satisfying Eq.~(\ref{4.2a5}) arises from the further calculation of $\tilde \xi _{}^{\left( 1 \right)}$.
This procedure requires solving the equation for $\tilde \xi _{}^{\left( 2 \right)}$, having the similar form to Eq.~(\ref{4.1a5}), where on the place of $q_0$ parameter, the $2q_0$ is.
However, this is a rather cumbersome procedure, and we only briefly describe it in the next section.
Here we only emphasize, that Fig.~\ref{fig:pic8a5} shows the region, more than sufficient to satisfy Eq.~(\ref{4.2a5}) and sufficient to satisfy the similar condition with $2q_0$ in place of $q_0$.
Applying the approximation Eq.~(\ref{4.2a5}) to Eq.~(\ref{4.1a5}), we transform it to the following form:
\begin{eqnarray}
2L{i_{{5 \mathord{\left/
				{\vphantom {5 2}} \right.
				\kern-\nulldelimiterspace} 2}}}\left( { - {e^\chi }} \right)\frac{{{\partial ^2}\tilde \varphi _1^{\left( 1 \right)}}}{{\partial {\chi ^2}}} + L{i_{{3 \mathord{\left/
				{\vphantom {3 2}} \right.
				\kern-\nulldelimiterspace} 2}}}\left( { - {e^\chi }} \right)\frac{{\partial \tilde \varphi _1^{\left( 1 \right)}}}{{\partial \chi }}
- L{i_{{1 \mathord{\left/
				{\vphantom {1 2}} \right.
				\kern-\nulldelimiterspace} 2}}}\left( { - {e^\chi }} \right)\tilde \varphi _1^{\left( 1 \right)} = 0.
\label{4.3a5}
\end{eqnarray}
Noticing the following polylogarithm property
\begin{eqnarray}
L{i_{s - 1}}\left( { - {e^\chi }} \right) = \frac{d}{{d\chi }}L{i_s}\left( { - {e^\chi }} \right),
\label{4.4a5}
\end{eqnarray}
we transform Eq.~(\ref{4.3a5}) to the form
\begin{eqnarray}
\frac{{{\partial ^2}}}{{\partial {\chi ^2}}}\left( {L{i_{\frac{5}{2}}}\left( { - {e^\chi }} \right)\tilde \varphi _1^{\left( 1 \right)}} \right) = \frac{3}{2}\frac{\partial }{{\partial \chi }}\left( {L{i_{\frac{3}{2}}}\left( { - {e^\chi }} \right)\tilde \varphi _1^{\left( 1 \right)}} \right).
\quad
\label{4.5a5}
\end{eqnarray}
By integrating the both sides of Eq.~(\ref{4.5a5}) and further applying simple transformations, it turns to the following form:
\begin{eqnarray}
L{i_{{5 \mathord{\left/
				{\vphantom {5 2}} \right.
				\kern-\nulldelimiterspace} 2}}}\left( { - {e^\chi }} \right)\frac{{\partial \tilde \varphi _1^{\left( 1 \right)}}}{{\partial \chi }} = \frac{1}{2}L{i_{{3 \mathord{\left/
				{\vphantom {3 2}} \right.
				\kern-\nulldelimiterspace} 2}}}\left( { - {e^\chi }} \right)\tilde \varphi _1^{\left( 1 \right)} + C,
\label{4.6a5}
\end{eqnarray}
where $C$ is the integration constant.
The last equation belongs to linear inhomogeneous type and it is solved by the method of the arbitrary constant variation.
In this case the solution of Eqs.~(\ref{4.3a5}),~(\ref{4.6a5}) is
\begin{eqnarray}
\tilde \varphi _1^{\left( 1 \right)}\left( \chi  \right) = C_1^{\left( 1 \right)}\sqrt { - L{i_{{5 \mathord{\left/
					{\vphantom {5 2}} \right.
					\kern-\nulldelimiterspace} 2}}}\left( { - {e^\chi }} \right)}
+ {C_2^{\left( 1 \right)}}\sqrt { - L{i_{{5 \mathord{\left/
					{\vphantom {5 2}} \right.
					\kern-\nulldelimiterspace} 2}}}\left( { - {e^\chi }} \right)} \int {d\chi } {\left( { - L{i_{{5 \mathord{\left/
						{\vphantom {5 2}} \right.
						\kern-\nulldelimiterspace} 2}}}\left( { - {e^\chi }} \right)} \right)^{ - \frac{3}{2}}}
,
\label{4.7a5}
\end{eqnarray}
where $C_1^{\left( 1 \right)}$, $C_2^{\left( 1 \right)}$ are the arbitrary integration constants.
The second partial solution in Eq.~(\ref{4.7a5}) increases in absolute value with increasing of $z$.
This fact leads to the first condition breaking in Eq.~(\ref{2.10a5}), starting from a certain value of $z$.
In this case the used perturbation theory becomes inapplicable and that is why $C_2^{\left( 1 \right)}$ should be set to zero.

So, we have the following expression for
$\tilde \varphi _1^{\left( 1 \right)}\left( z \right)$:
\begin{eqnarray}
\tilde \varphi _1^{\left( 1 \right)}\left( \chi  \right) = C_1^{\left( 1 \right)}\sqrt { - L{i_{{5 \mathord{\left/
					{\vphantom {5 2}} \right.
					\kern-\nulldelimiterspace} 2}}}\left( { - {e^\chi }} \right)}.
\label{4.8a5}
\end{eqnarray}
The general solutions of the second and the third equations in Eq.~(\ref{2.20a5}) have the form:
\begin{eqnarray}
\tilde \varphi _2^{\left( 1 \right)}(z) = C_1^{(2)}{e^{{q_0}z}} + C_2^{(2)}{e^{ - {q_0}z}},
\quad
\tilde \varphi _3^{\left( 1 \right)}(z) = C_1^{(3)}{e^{{q_0}z}} + C_2^{(3)}{e^{ - {q_0}z}}.
\label{4.9a5}
\end{eqnarray}
Taking into account the finiteness of electric field value at $z \to  - \infty $, $C_2^{\left( 3 \right)}$ constant in Eq.~(\ref{4.9a5}) should be set to zero, $C_2^{\left( 3 \right)} \equiv 0$.
Arbitrary constants $C_1^{(1)}$, $C_1^{(2)}$, $C_2^{(2)}$ and $C_1^{(3)}$ can be obtained from the boundary conditions in Eq.~(\ref{2.21a5}).
It is easily seen, that these constants are linear in ${\tilde \xi ^{\left( 1 \right)}}$.
According to Eqs.~(\ref{2.11a5}),~(\ref{2.19a5}), and~(\ref{2.20a5}), ${\tilde \xi ^{\left( 1 \right)}}$ is the first harmonic of the Fourier transform of spatially periodic profile perturbation of liquid dielectric film surface.
\begin{eqnarray}
\nonumber
\tilde \varphi _1^{\left( 1 \right)}\left( \chi  \right) = {E_0}{\tilde \xi ^{\left( 1 \right)}}\sqrt {{{L{i_{{5 \mathord{\left/
							{\vphantom {5 2}} \right.
							\kern-\nulldelimiterspace} 2}}}\left( { - {e^\chi }} \right)} \mathord{\left/
			{\vphantom {{L{i_{{5 \mathord{\left/
										{\vphantom {5 2}} \right.
										\kern-\nulldelimiterspace} 2}}}\left( { - {e^\chi }} \right)} {L{i_{{5 \mathord{\left/
										{\vphantom {5 2}} \right.
										\kern-\nulldelimiterspace} 2}}}\left( { - {e^{{\chi _0}}}} \right)}}} \right.
			\kern-\nulldelimiterspace} {L{i_{{5 \mathord{\left/
							{\vphantom {5 2}} \right.
							\kern-\nulldelimiterspace} 2}}}\left( { - {e^{{\chi _0}}}} \right)}}} G\left( {{q_0}} \right),
\quad
\tilde \varphi _2^{\left( 1 \right)}\left( z \right) = {\tilde \xi ^{\left( 1 \right)}}\left( {{e^{q\left( {z - \bar \xi } \right)}} - C{e^{q\left( {\bar \xi  - z} \right)}}} \right){E_0}F\left( q \right),
\\
\tilde \varphi _3^{\left( 1 \right)}\left( z \right) = {\tilde \xi ^{\left( 1 \right)}}{e^{q\left( {z - \bar \xi } \right)}}{E_0}F\left( q \right)\left( {1 - \delta } \right),
\quad
\label{4.8aa5}
\end{eqnarray}
where the following notations are introduced:
\begin{eqnarray}
\nonumber
G\left( {{q_0}} \right) = \frac{{L{i_{\frac{5}{2}}}\left( { - {e^{{\chi _0}}}} \right)\left( { 2 b{z_0}\frac{n}{{{n_s}}} + {y_0}\left( {\varepsilon  - 1} \right)} \right)}}{{L{i_{\frac{5}{2}}}\left( { - {e^{{\chi _0}}}} \right)\varepsilon {y_0} + b L{i_{\frac{3}{2}}}\left( { - {e^{{\chi _0}}}} \right)}},
\quad
F\left( {{q_0}} \right) = \frac{{\left( {\frac{1}{\varepsilon } - 1} \right)L{i_{\frac{3}{2}}}\left( { - {e^{{\chi _0}}}} \right) + L{i_{\frac{5}{2}}}\left( { - {e^{{\chi _0}}}} \right)2{z_0}\frac{n}{{{n_s}}}}}{{\left( {1 + C} \right)\left( {L{i_{\frac{5}{2}}}\left( { - {e^{{\chi _0}}}} \right)\varepsilon {y_0} + b L{i_{\frac{3}{2}}}\left( { - {e^{{\chi _0}}}} \right)} \right)}},
\\
{y_0} = 2{q_0}{z_0},
\quad
b  = {{\left( {1 - C} \right)} \mathord{\left/
		{\vphantom {{\left( {1 - C} \right)} {\left( {1 + C} \right)}}} \right.
		\kern-\nulldelimiterspace} {\left( {1 + C} \right)}},
\quad
\delta  \equiv {{\left( {{\varepsilon _d} - \varepsilon } \right)} \mathord{\left/
		{\vphantom {{\left( {{\varepsilon _d} - \varepsilon } \right)} {\left( {{\varepsilon _d} + \varepsilon } \right)}}} \right.
		\kern-\nulldelimiterspace} {\left( {{\varepsilon _d} + \varepsilon } \right)}},
\quad
C = \delta {e^{ - 2{q_0}\left( {d + \bar \xi } \right)}}.
\quad
\label{4.9aa5}
\end{eqnarray}

The general solution of Eq.~(\ref{2.21a5}) in the analogous approximation for $\tilde \varphi _j^{\left( e \right)\left( 1 \right)}$, $j = 1,2,3$ has the form:
\begin{eqnarray}
\nonumber
\tilde \varphi _1^{\left( e \right)}\left( z \right) = C_2^{\left( {e1} \right)}{e^{ - {q_0}z}},
\quad
\tilde \varphi _3^{\left( e \right)\left( 1 \right)}\left( z \right) = C_1^{(e3)}{e^{{q_0}z}},
\quad
\tilde \varphi _2^{\left( e \right)\left( 1 \right)}\left( z \right) = C_1^{\left( {e2} \right)}{e^{{q_0}z}} + C_2^{\left( {e2} \right)}{e^{ - {q_0}z}}.
\label{4.10a5}
\end{eqnarray}
The form of Eq.~(\ref{4.10a5}) is chosen to satisfy the finiteness conditions  Eq.~(\ref{2.6a5}) at $z \to  \pm \infty $.
The constants in Eq.~(\ref{4.10a5}) can also be obtained from the corresponding linear approximation of boundary conditions in Eq.~(\ref{2.21a5}).
So, the expression for $\tilde \varphi _j^{\left( e \right)\left( 1 \right)}$ has the form:
\begin{eqnarray}
\nonumber
\tilde \varphi _1^{\left( e \right)\left( 1 \right)}\left( z \right) = \left( {\varepsilon  + 1} \right)CE{\tilde \xi ^{\left( 1 \right)}}{F^{\left( e \right)}}\left( q \right){e^{q\left( {\bar \xi  - z} \right)}},
\quad
\tilde \varphi _2^{\left( e \right)\left( 1 \right)}\left( z \right) = {\tilde \xi ^{\left( 1 \right)}}E{F^{\left( e \right)}}\left( {{q_0}} \right)\left( {C{e^{{q_0}\left( {\bar \xi  - z} \right)}} - {e^{{q_0}\left( {z - \bar \xi } \right)}}} \right),
\\
\tilde \varphi _3^{\left( e \right)\left( 1 \right)}\left( z \right) = \left( {1 - \delta } \right){\tilde \xi ^{\left( 1 \right)}}E{F^{\left( e \right)}}\left( {{q_0}} \right){e^{{q_0}\left( {z - \bar \xi } \right)}},
\quad
{F^{\left( e \right)}}\left( q \right) = {{\left( {1 - {\varepsilon ^{ - 1}}} \right)} \mathord{\left/
		{\vphantom {{\left( {1 - {\varepsilon ^{ - 1}}} \right)} {\left( {\varepsilon \left( {1 + C} \right) + 1 - C} \right)}}} \right.
		\kern-\nulldelimiterspace} {\left( {\varepsilon \left( {1 + C} \right) + 1 - C} \right)}}.
\quad
\label{4.11a5}
\end{eqnarray}

Let us consider the last equation in Eq.~(\ref{2.21a5}).
Taking into account Eqs.~(\ref{4.8aa5}),~(\ref{4.11a5}),~(\ref{2.12a5}), it can be written in the following form:
\begin{eqnarray}
\Phi \left( {{q_0}} \right){\tilde \xi ^{\left( 1 \right)}} = 0,
\label{4.12a5}
\end{eqnarray}
where we introduce the following function
\begin{eqnarray}
\Phi \left( {{q_0}} \right) \equiv \frac{{4\pi \alpha }}{{{{E_0^2}}}}\left( {{\kappa ^2} + q_0^2\beta } \right) - \frac{n}{{{n_s}}}\left( {1 - G\left( {{q_0}} \right)} \right)
- \frac{n}{{{n_s}}}\left( {1 + C} \right){y_0}\left( {F\left( {{q_0}} \right) + {\frac{{{F^{\left( e \right)}}\left( {{q_0}} \right)}}{4}}} \right),
\quad
\beta  = 1 + \frac{{{\kappa ^2}{{\bar \xi }^2}}}{2}.
\label{4.13a5}
\end{eqnarray}

\begin{figure}
	 \begin{minipage}{.5\textwidth}
	 	\includegraphics[scale=0.85]{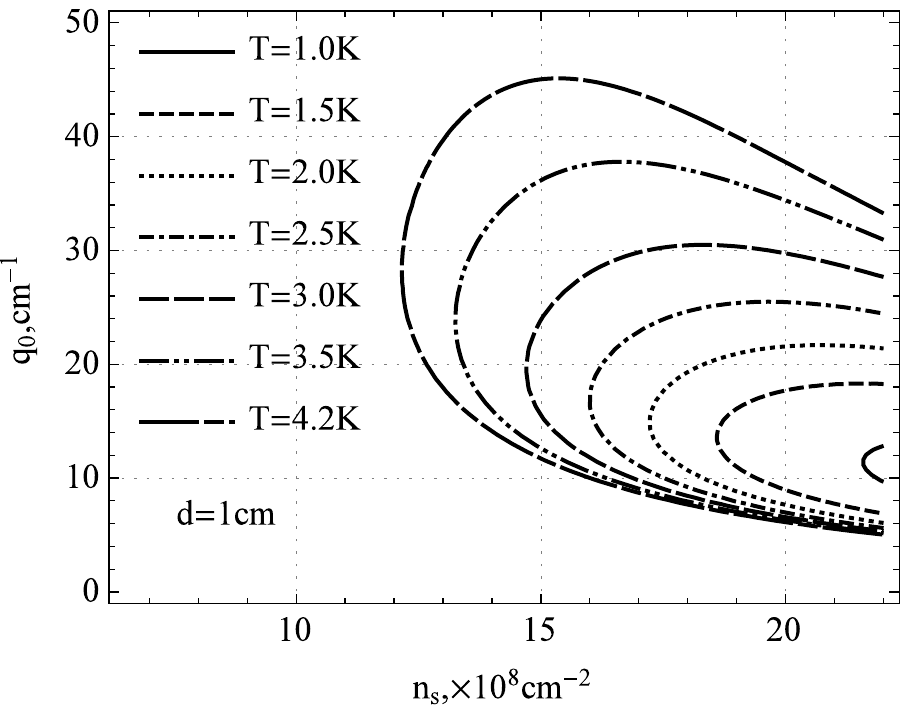}
	 	\caption{\label{fig:pic9aa5}
	 		Critical curves ${q_0}\left( {{n_{sc}}} \right)$ in massive helium case.}
	 \end{minipage}
	\hfill
	 \begin{minipage}{.5\textwidth}
	 \includegraphics[scale=0.85]{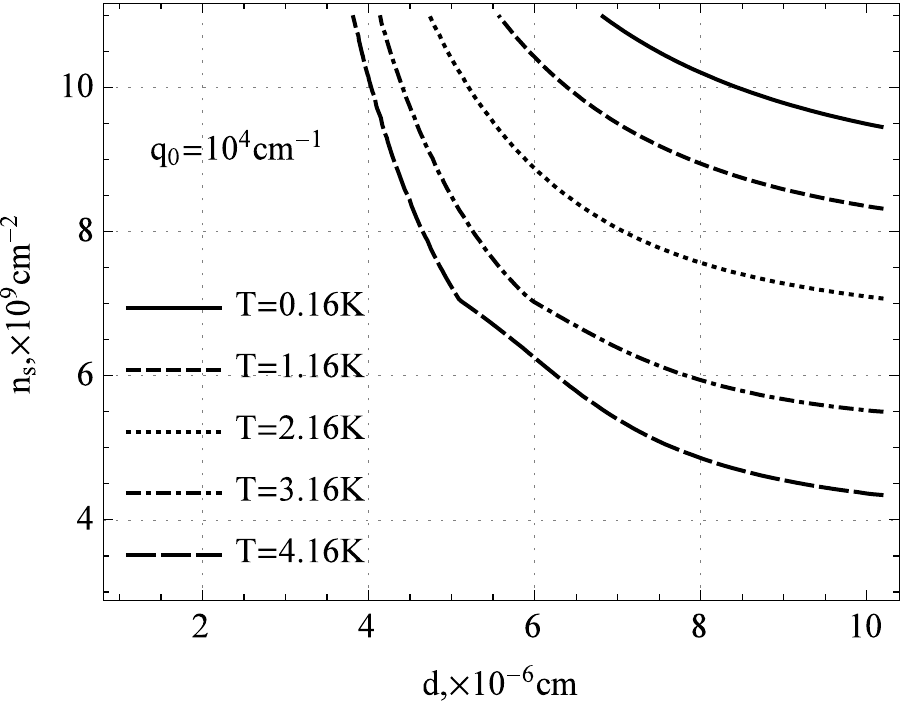}
	 \caption{\label{fig:pic9ba5}
	 	Critical curves ${n_{sc}}\left( d \right)$ in thin helium film case.}
	 \end{minipage}
\end{figure}

It is easily seen, that Eq.~(\ref{4.12a5}) has two solutions: ${\tilde \xi ^{\left( 1 \right)}} = 0$ and $\Phi \left( {{q_0}} \right) = 0$.
The first trivial solution ${\tilde \xi ^{\left( 1 \right)}} = 0$ describes the absence of spatially periodic structures on liquid dielectric surface.
So, in this case the liquid dielectric surface remains flat.
In the case of the phase transition to the state with spatially periodic profile of liquid dielectric surface, we consider ${\tilde \xi ^{\left( 1 \right)}} \ne 0$ (see Eqs.~(\ref{2.7a5}),~(\ref{2.11a5}),~(\ref{2.12a5})).
So, in this case we choose the second solution
\begin{eqnarray}
\Phi \left( {{q_0}} \right) = 0.
\label{4.14a5}
\end{eqnarray}
According to Eq.~(\ref{4.9aa5}), the last equation defines the value of reciprocal lattice vector as a function of the phase transition parameters: temperature $T_c$, external clamping field $E_c$ (or electrons areal density $n_{sc}$, see Eq.~(\ref{3.12ba5})), dielectric density $\rho $, its surface tension $\alpha $ and permittivity $\varepsilon $, and the solid substrate permittivity ${\varepsilon _d}$.
In fact, Eq.~(\ref{4.14a5}) defines a certain critical surface ${q_0} = {q_0}\left( {{n_{sc}},{T_c}} \right)$ of the phase transition.
The procedure of such surface obtaining goes out of analytical approach and requires numeric calculations.
As the result we obtain the dependency, presented on Figs.~\ref{fig:pic9aa5}~and~\ref{fig:pic9ba5}.

Fig.~\ref{fig:pic9aa5} shows curves ${q_0} = {q_0}\left( {{n_{sc}}} \right)$ for different fixed temperatures and takes into account Eq.~(\ref{3.13da5}) in the case of liquid helium film, having thickness $d = 1cm$.
This data is obtained from Eq.~(\ref{4.14a5}) and it is in good agreement with the experimental data~\cite{pla1979LeidererW,ss1982LeidererES}.
According to Fig.~\ref{fig:pic9aa5}, at $T = 3,5K$ the periodic structures, having reciprocal lattice period ${q_0} \approx 27c{m^{ - 1}}$ (corresponding to lattice distance $a = 2\pi q_0^{ - 1} \approx 0.28cm$), start appearing at ${n_{sc}} > 1,2 \cdot {10^9}c{m^{ - 2}}$.
From the other hand, according to Ref.~\cite{ss1982LeidererES}, at $T= 3,5K$ the lattice period distance is $a = 0.24cm$.
Again, according to Ref.~\cite{ss1982LeidererES}, at $T = 2,5K$ the periodic structures appear at the clamping field values, higher than  ${E_c} = 2600{V \mathord{\left/
		{\vphantom {V {cm}}} \right.
		\kern-\nulldelimiterspace} {cm}} = 4\pi e \cdot 1,38 \cdot {10^9}c{m^{ - 2}}$.
In our case, as seen from Fig.~\ref{fig:pic9aa5}, at $T = 2,5K$ the critical value of external clamping field is ${E_c} \approx 4\pi e \cdot 1,6 \cdot {10^9}c{m^{ - 2}}$.
According to Ref.~\cite{pla1979LeidererW}, at $T = 4,2K$, the lattice period distance is $a = 0.176cm$, which is forming at external clamping field values, higher than ${E_c} \approx 4\pi e \cdot 0,95 \cdot {10^9}c{m^{ - 2}}$.
And according to the present paper calculus (see Fig.~\ref{fig:pic9aa5}), the corresponding values are: $T = 4,2K$, $a = 2\pi q_0^{ - 1} \approx 0.22cm$ and ${E_c} \approx 4\pi e \cdot 1,22 \cdot {10^9}c{m^{ - 2}}$.

Let us emphasize, that Refs.~\cite{pla1979LeidererW,ss1982LeidererES} described not only the case of quasi-neutrality of the system (see Eq.~(\ref{3.12ba5})), but the case of charged systems too.
For this reason it is convenient to relate the forming of spatially periodic structures to the critical value of clamping field ${E_c}$, but not to ${n_{sc}}$.
The value of ${n_{sc}}$ defines the settlement rate of dielectric surface with dimples.
In other words, it defines the number of dimples per unit of dielectric surface area.
Indeed, in the case of small ${n_s}$ values, if the clamping field exceeds the value of $E_c$, it is possible to observe only several dimples (e.g., 2,8,20).
With the increasing of $n_s$ value, the number of dimples increases too and the helium surface becomes more densely filled with the dimples.
On reaching the value of $n_s^m \approx 2 \cdot {10^9}c{m^{ - 2}}$ the considered periodic structure starts breaking, because the electron clusters from  dimples start sinking in helium in the form of bubbles~\cite{el1987AlbrechtL}, that move towards metal substrate electrode, generating the clamping field.
In the case of ${n_s} < n_s^m$, the periodic structures can also be formed in quasi-neutral system, if  $n_s^m > {n_s} > {E_c}{\left( {4\pi e} \right)^{ - 1}}$.
Further on we don't consider the forming of periodic dimple structures in charged system.
In general case this problem has a separate solution.
And for the non-degenerate gas of charges above liquid dielectric surface this problem was solved in Ref.~\cite{jmp2012LytvynenkoSK}.
An important property of ``massive'' helium is the dependence of ${E_c}$ (or ${n_{sc}}$ in the quasi-neutral system case, see Eq.~(\ref{3.12ba5})) on the temperature.
As in experiments~\cite{pla1980EbnerL}, in our case $E_c$ decreases with $T$ growth (see Fig.~\ref{fig:pic9aa5}).

Now, let us consider the phase transition in the case of thin film of liquid dielectric.
In this situation the Van der Waals forces, acting on the dielectric atom, predominate over the gravitation ones.
In such system the possibility of theoretical prediction and experimental registration of some interesting effects arises.
As seen on Figs.~\ref{fig:pic4aa5},~\ref{fig:pic4ba5}, thin dielectric films provide the several orders higher values of permitted ${n_s}$, than the massive ones do.
The numeric estimates show, that the quite acceptable $n_s$ values for the phase transition observation are  ${n_s} \sim {10^{12}}{cm^{ - 2}}$ or even higher.
It was shown in Ref.~\cite{prb1990HuD}, that the helium films, having thickness of several hundred angstroms, are stable for all $n_s$ values.
And only by diminishing the film thickness up to $50$ angstroms or less, the electrons begin tunneling through the film towards the solid substrate.
In the theoretical Ref.~\cite{prl1983PeetersP} it was shown, that for helium film with $d=100 {{\AA}}$,  located on metallic substrate, Wigner crystal was formed in low temperature range at ${n_{sc1}} \approx {10^{11}}{cm^{ - 2}}$ or higher.
The further increasing of $n_s$ value up to ${n_{sc2}} \approx 1,37 \cdot {10^{12}}{cm^{ - 2}}$ leads to the so-called crystal quantum melting effect.
The similar to the quantum melting effect was registered in Ref.~\cite{ss1996GunslerBMNL}, however, at lower ${n_{sc2}}$ value, than Ref.~\cite{prl1983PeetersP} predicted.

In terms of present paper approach, the numeric evaluations of Eq.~(\ref{4.14a5}) at $T = 0,1K$ and $d = {10^{ - 6}}cm$ also show, that in the range ${n_{sc1}} \le {n_s} \le {n_{sc2}}$ the periodic structures can exist in large scope of ${q_0}$ values.
The limit values ${n_{sc1}} \approx {10^{11}}{cm^{ - 2}}$ and ${n_{sc2}} \approx 2,4 \cdot {10^{12}}{cm^{ - 2}}$ can be interpreted as the points of the structures appearing and disappearing (melting) correspondingly.
The obtained upper limit value ${n_{sc2}}$ significantly differs from the analogous value in Ref.~\cite{prl1983PeetersP}.
However, in the range of high density and low temperature values our approach  requires more rigorous clarification, as in this range a significant role can play quantum effects, e.g., the exchange interaction.
For this reason the given estimates demonstrate only qualitative agreement with the similar effect evaluations in Ref.~\cite{prl1983PeetersP} and cannot provide a valid quantitative agreement.
In the end of this section let us also notice the qualitative agreement between the ${n_{sc}}\left( d \right)$ dependence (see Fig.~\ref{fig:pic9ba5}), obtained basing on Eqs.~(\ref{4.13a5}),~(\ref{4.14a5}), and the experimental data~\cite{prb1997MisturaGNL}.

\section{The order parameter of the phase transition to the symmetric phase near the critical point}

Before starting the calculation of amplitude of the studied spatially periodic structures, let us make the following methodological note.
As mentioned before, the forming of spatially periodic structures (considering them two-dimensional!) in the system of charges above liquid dielectric surface is called Wigner crystallization~\cite{prl1979GrimesA}.
On the other hand, dimple crystals are associated with charges, located in dimples of liquid dielectric surface, forming a periodic structure~\cite{pla1979LeidererW}.
It is obvious, that the dimple structure is three-dimensional.
In present paper approach the considered periodic structures are associated with spatially periodic deformation of liquid dielectric surface.
In other words, all the described here periodic structures can be considered as dimple type ones.
So, the proposed approach seems to be available only for dimple crystals description, but not for Wigner crystal description.
However, the depth of the dimples can vary in value.
In the case of extremely shallow dimples, their small depth doesn't affect the experimental data, concerned with the periodic structures existence.
So, from the experimental point of view, these structures are perceived to be two-dimensional.
Thus, the approach to considering these periodic structures as two-dimensional or three-dimensional, depends on their properties (e.g., dimple depth) and experimental registration technique.
In macroscopic dimple case, the dimples are visible.
And in the Wigner crystallization case, the phase transition is registered by indirect parameters, concerned with dynamical properties of the system~\cite{prl1979GrimesA}, while the dielectric surface is assumed to be plane.
In the last example we may deal with the case of small deformation of dielectric surface, making no affect on the experimental measurement process.

To make a theoretical grounding for this assumption let us obtain the value of the order parameter $\tilde \xi _{}^{\left( 1 \right)}$ near the phase transition point.
This is a cumbersome procedure, so, in this paper we do not present the detailed calculations.
We only briefly describe the procedure and demonstrate the main results.
In Ref.~\cite{jps2015SlyusarenkoL} the procedure of obtaining the amplitude of spatially periodic states was considered in more detailed way for the case of non-degenerate gas.

To obtain $\tilde \xi _{}^{\left( 1 \right)}$ we have to make expansion of Eqs.~(\ref{1.1a5}),~(\ref{1.4a5}) on small perturbations $\tilde \xi \left( {\boldsymbol{\rho }} \right)$, ${\tilde \varphi _j}\left( {z,{\boldsymbol{\rho }}} \right)$, $\tilde \varphi _j^{\left( e \right)}\left( {z,{\boldsymbol{\rho }}} \right)$ and small differences $T - {T_c}$, $E - {E_c}$ (or $n_s - {n_{sc}}$ in quasi-neutrality case Eq.~(\ref{3.12ba5})).
Then, making the Fourier transforms Eqs.~(\ref{2.11a5}),~(\ref{2.13a5}) and taking into account the main approximation Eqs.~(\ref{2.14a5})~-~(\ref{2.18a5}) and linear approximation Eqs.~(\ref{2.21a5})~-~(\ref{2.23a5}) of the considered perturbation theory, we obtain the next non-vanishing approximation at $q = q_0^{}$.
On making these calculations it becomes obvious, that we also have to obtain the relation between the first $\tilde \xi _{}^{\left( 1 \right)}$ and the second $\tilde \xi _{}^{\left( 2 \right)}$ harmonics.
For this reason we have to expand Eqs.~(\ref{1.1a5}),~(\ref{1.4a5}) on small perturbations $\tilde \xi \left( {\boldsymbol{\rho }} \right)$, ${\tilde \varphi _j}\left( {z,{\boldsymbol{\rho }}} \right)$, $\tilde \varphi _j^{\left( e \right)}\left( {z,{\boldsymbol{\rho }}} \right)$ and small differences $T - {T_c}$, $E - {E_c}$.
And then take the Fourier transform of this expansion at $q = 2q_0^{}$.
Basing on the methods, developed in the previous section and Ref.~\cite{jps2015SlyusarenkoL}, we obtain the relation between the first and second harmonics of the Fourier transform of the order parameter $\tilde \xi$:
\begin{eqnarray}
\tilde \xi _{}^{\left( 2 \right)} = \frac{\gamma }{{{z_0}}}{\left( {\tilde \xi _{}^{\left( 1 \right)}} \right)^2},
\label{4.27a5}
\end{eqnarray}
that leads to the non-linear equation for the amplitude ${\tilde \xi ^{\left( 1 \right)}}$ obtaining:
\begin{eqnarray}
{\left( {{{\tilde \xi }^{\left( 1 \right)}}} \right)^3} = {\tilde \xi ^{\left( 1 \right)}}\frac{{z_0^2\Psi \left( {{E_c},{T_c}} \right)}}{n}\left(\frac{{\partial n}}{{\partial E}}\left( {E - {E_c}} \right)
+\frac{{\partial n}}{{\partial T}}\left( {T - {T_c}} \right)\right).
\label{4.28a5}
\end{eqnarray}
We do not present here the explicit expressions for $\gamma$ and $\Psi \left( {{E_c},{T_c}} \right)$ functions, because of their cumbersome structure and complicated dependence on ${E_c}$ and ${T_c}$.
Eq.~(\ref{4.28a5}) has two solutions.
The first solution ${\tilde \xi ^{\left( 1 \right)}} = 0$ is trivial and it does not describe any phase transition.
So, for the same reason as on obtaining the critical curve (see Eqs.~(\ref{4.12a5}~-~(\ref{4.14a5}))), we do not consider this solution.
Further on, during the numeric evaluations of the amplitude ${\tilde \xi ^{\left( 1 \right)}}$ value at fixed $E_c$ and $T_c$ values, we use the explicit expressions for $\gamma$ and $\Psi \left( {{E_c},{T_c}} \right)$.
So as in the case of calculating the first harmonic of the Fourier transform of the density perturbation at $z = \bar \xi$ (see Eqs.~(\ref{3.6a5}),~(\ref{4.1a5}),~(\ref{4.8a5}):
\begin{eqnarray}
\nonumber
{n^{\left( 1 \right)}} =  - T\frac{{\partial n}}{{\partial \mu }}\left( {1 - G\left( {{q_0}} \right)} \right)\frac{{{{\tilde \xi }^{\left( 1 \right)}}}}{{{z_0}}}.
\label{4.29a5}
\end{eqnarray}
The numeric estimates show, that $G\left( {{q_0}} \right) < 1$ (see Eq.~(\ref{4.9a5})).
According to Eq.~(\ref{4.29a5}), in the case ${\tilde \xi ^{\left( 1 \right)}} \ne 0$ we have the situation, when above the dimples on liquid dielectric surface the charge density maximums are located.
Also, above the ``hills'' on dielectric surface the charge density minimums are located.
We have a similar situation in the case of the second solution of Eq.~(\ref{4.28a5})
\begin{eqnarray}
\nonumber
{\tilde \xi ^{\left( 1 \right)}} = z_0^{}\sqrt {\Psi \left( {\frac{{\partial \ln n}}{{\partial E}}\left( {E - {E_c}} \right) + \frac{{\partial \ln n}}{{\partial T}}\left( {T - {T_c}} \right)} \right)}.
\end{eqnarray}
Evaluating this expression at $T = 2,5K$, ${n_s} = 1,4 \cdot {10^9}c{m^{ - 2}}$, $d = 0.1cm$ and ${q_0} = 23c{m^{ - 1}}$ parameter values and $\sqrt {\left( {{E \mathord{\left/
				{\vphantom {E {{E_c}}}} \right.
				\kern-\nulldelimiterspace} {{E_c}}}} \right) - 1}  \approx 0.1$, $\sqrt {\left( {{{{T_c}} \mathord{\left/
				{\vphantom {{{T_c}} T}} \right.
				\kern-\nulldelimiterspace} T}} \right) - 1}  \approx 0.1$, we obtain ${\tilde \xi ^{\left( 1 \right)}} \approx 4,7 \cdot {10^{ - 11}}cm$.
This value is in satisfactory agreement with the corresponding estimations of this quantity in Refs.~\cite{book1997Andrei, book2003MonarkhaK}.
This estimation can hardly have physical interpretation,  because its value is several orders less than typical atom size ${a_0} \sim {10^{ - 8}}cm$.
For this reason the dielectric surface can be considered plane.
However, our estimations are made in the region ${E \mathord{\left/
		{\vphantom {E {{E_c}}}} \right.
		\kern-\nulldelimiterspace} {{E_c}}} \sim 1$, ${T \mathord{\left/
		{\vphantom {T {{T_c}}}} \right.
		\kern-\nulldelimiterspace} {{T_c}}} \sim 1$, where the proposed perturbation theory takes place.
Moving far from the transition point (decreasing temperature, increasing of clamping field) can significantly change the surface structure, including the 3D period structure appearing.

Thus, for the purpose of simplicity we demonstrate the benefits of the developed approach by studying the phase transition to the structured state, characterized by a single reciprocal lattice distance ${q_0}$.
However, under the conditions of real experiment~\cite{pla1980EbnerL}, the periodic wavy structure on the dielectric surface is observed as intermediate state during the phase transition from the homogeneous state to the 2D hexagonal structure.
The description of periodic structures, characterized by two independent reciprocal lattice vectors in the parallel plane to the dielectric surface, is a separate problem, waiting for the solution in terms of the developed approach.

\section{Conclusion}

Summarizing the present paper, we develop quantum-statistic theory of equilibrium spatially inhomogeneous states of the system of charges above liquid dielectric surface in external clamping electric field.
The state of the system is considered to be quasi-neutral, i.e., the field, induced by charges, compensates the external electric field at infinity.
The theory is developed in quasi-classical approach, applying the concept of Wigner distribution function of electrons above liquid dielectric surface.

Beyond the scope of Boltzmann statistics we obtain the self-consistency equations, describing the phase transition in the system to the state with spatially periodic structures near the critical point.
The benefits of this approach are demonstrated by describing the phase transition with the forming of spatially periodic structures of wave type.
Applying the analytical and numeric methods, we analyze the influence of dielectric film thickness on the critical parameters of the studied phase transition.
We discuss the criterion on the system stability against the possible quantum tunneling of electrons to the solid substrate.
The obtained results are compared to the theoretical end experimental data, previously obtained by other authors.

At least, the present approach requires modification in two directions.
Firstly, this approach can be generalized to the description of spatially periodic structures, characterized by two independent reciprocal (and direct too) lattice vectors.
Secondly, the theory can be modified to take into account the effect of quantum effects, such as exchange interaction.
In present time the authors are working on both problems.

%\section*{Acknowledgment}
%Authors want to thank V.N.Karazin Kharkiv Natiational University and NSC KIPT for the support in writing this paper.

\section*{References}

%\bibliography{bibliography}

\end{document}